\documentclass[aps,physrev,preprint,groupedaddress]{revtex4-2}

\usepackage{physics}
\usepackage{amsmath}
\usepackage{xcolor}
\usepackage{graphicx}

\begin{document}


\title{Quantized non-Abelian helicity of flat bands in 2D Floquet topological photonic insulators}


\author{Bo Leng and Vien Van}
\affiliation{Department of Electrical and Computer Engineering, University of Alberta, Edmonton, Alberta, Canada, T6G 2V4}



\begin{abstract}
Flat-band states in topological systems provide a unique platform for investigating strongly correlated phenomena and many body physics.
However, in 2D static tight-binding systems, perfectly flat bands can only exist in the topologically trivial phase, as characterized by a zero Chern number.  
Here we show that by introducing periodic driving into a 2D photonic Lieb lattice composed of coupled microring resonators, the resulting Floquet topological insulator can host perfectly flat bands with nontrivial topology.
In particular, by tracking the evolution of the flat-band modes over each cycle, we show that the non-Abelian displacements of the flat-band modes are characterized by a nontrivial quantized helicity even though the quasi-energy bands have zero Chern number.  The helical motion of the flat-band modes can be described by a braiding of the world lines of their trajectories, with a nontrivial winding number directly connected to the helicity.  We also propose a scheme to experimentally measure the quantized non-Abelian helicity in a microring lattice subject to a synthetic magnetic field.  These results suggest that Floquet topological photonic insulators based on coupled microring resonators can provide a versatile platform for investigating non-Abelian topological physics and strongly correlated phenomena in photonic flat-band systems.
\end{abstract}


\maketitle

\section{Introduction}
Topological insulators exhibit many unique properties which have attracted much recent interest, both for investigating fundamental topological physics as well as for exploring potential applications in quantum information.
Among the many novel manifestations of topological insulators are systems with non-Abelian gauge structures, which typically arise in quantum systems possessing degenerate eigenstate subspaces.
The evolution of these systems can be described by braid structures, which exhibit anyonic statistics and are the cornerstone of holonomic quantum computing.  Motivated by this, various models and systems exhibiting non-Abelian topological behavior have been proposed and investigated, including cold atoms \cite{Ruseckas2005, Sugawa2018, Sugawa2021}, exciton polariton systems \cite{Tercas2014,  Polimeno2021}, photonics \cite{Bliokh2007, iadecola2016non, Chen2019, Yang2019, zhang2022non, Cheng2023PRL, cheng2025non}, and electrical and superconducting circuits \cite{Wu2022, Abdumalikov2013}.  Among these, photonics represents a particular promising platform owing to the flexibility in designing photonic lattices to emulate a specific Hamiltonian.  Non-Abelian photonic systems have been experimentally demonstrated using coupled waveguide arrays \cite{noh2020braiding, zhang2022non, sun2022non}, microring lattices \cite{Cheng2023PRL}, and fiber optics \cite{Yang2019, cheng2025non}.  For example, non-Abelian gauge structure in a coupled waveguide system has been shown to generate different geometric phases for different permutations of the coupling sequence \cite{zhang2022non}. Non-Abelian braiding has also been measured in topological photonic Majorana zero mode lattices \cite{noh2020braiding}.  
One-dimensional coupled waveguide arrays supporting degenerate flat bands have been shown to exhibit different light paths for different non-Abelian Thouless pumping sequences \cite{brosco2021non, sun2022non}. Non-Abelian behaviors have also been demonstrated in 2D photonic lattices with synthetic dimensions using ring resonators \cite{Cheng2023PRL,cheng2025non}.

A common feature of these non-Abelian photonic systems is that their behavior is described by an adiabatic evolution process which takes place in a degenerate eigenstate subspace characterized by matrix-valued connections.
Non-Abelian behavior can also arise in Floquet topological photonic insulators (TPIs) with non-adiabatic periodic driving even though the quasienergy bands are non-degenerate. The non-adiabatic driving causes the bands to mix so that the evolution of the system must take into account the micromotions of all the Floquet states.
An example of such a system which has been experimentally demonstrated \cite{Afzal2020} is a Floquet insulator based on a 2D coupled microring lattice, in which the uni-directional circulation of light in each microring emulates a time-like dimension. 
The lattice effectively behaves as a periodically driven system, which has been shown to exhibit topological behaviors not observed in static or adiabatic driving systems, such as the anomalous Floquet insulator phase \cite{Afzal2020}.  
When arranged in a 2D Lieb configuration, the resulting Floquet Lieb insulator (FLI) hosts a symmetry-protected flat band at zero quasienergy, with two dispersive bulk bands that also become flattened under perfect coupling condition \cite{Song2024_Commun}. Such an all-bands-flat (ABF) system can provide an interesting platform for investigating strongly-correlated phenomena such as Anderson localization \cite{Goda2006, Chalker2010}, quantum caging \cite{Danieli2021b} and the fractional quantum Hall effect \cite{Tang2011, Wang2011}.  However, the flat bands of the FLI lattice have trivial Chern number, even though they are embedded in band gaps with nontrivial winding numbers.  On the other hand, it has been shown that in a 2D static system with finite hopping range, a perfectly flat band must have zero Chern number \cite{Kruchkov2022}.  This raises the question whether flat bands in a periodically-driven 2D system can be classified by a nontrivial topological invariant.

In this paper we investigate the topological properties of the flat-band modes in a 2D Floquet microring lattice by tracking the non-Abelian evolution of the modes over each cycle.
We show that while the Wannier centers of the flat-band modes exhibit zero displacement, the non-Abelian displacements are nontrivial, exhibiting helical micro-motions which are characterized by a nontrivial quantized helicity.
This quantized non-Abelian helicity is topological in origin and can be used to classify perfectly flat bands in Floquet systems.
We also propose a scheme to experimentally measure the quantized helicity of a microring lattice with all flat bands.  Finally, we show that the evolution of the flat-band modes in real space can be represented by world lines whose braiding structure is characterized by a nontrivial winding number that is connected to the non-Abelian helicity. 
Our results suggest that Floquet topological insulators based on microring lattices could serve as a versatile platform for investigating non-Abelian physics and novel phenomena associated with topological flat-band systems.

\begin{figure}[t]
    \centering\includegraphics[scale=0.48]{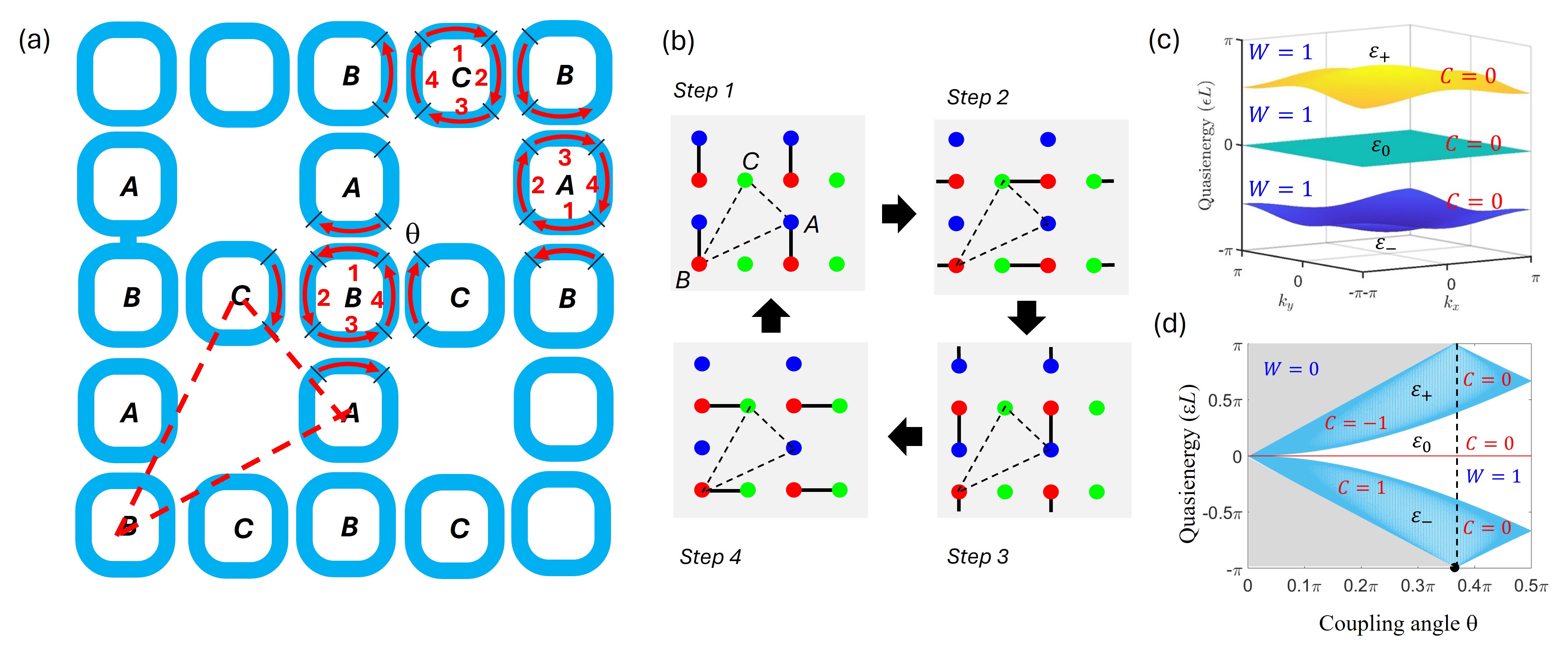}
    \caption{(a) Schematic of the Floquet-Lieb insulator realized using coupled microring resonators.  The dashed red lines indicate a unit cell consisting of three microrings A, B, C.  The arrows in the rings show light circulating around each ring and coupling to neighbor resonators. (b) Coupling sequence of the FLI lattice with each dot representing a site microring waveguide.  (c) Quasienergy band diagram of the FLI with coupling angle $\theta = 0.45 \pi$. The Chern numbers ($C$) of the bands and winding numbers ($W$) of the band gaps are also shown.  (d) Quasienergy spectrum of the FLI (blue color) projected over all $(k_x, k_y)$ values vs. the coupling angle.  Grey regions indicate trivial bandgaps with winding number $W = 0$; white regions indicate nontrivial bandgaps with $W = 1$.}
    \label{fig:FLI}
\end{figure}

\section{Floquet-Lieb microring lattice}

We consider a Floquet-Lieb insulator constructed using a 2D depleted square lattice of coupled microring resonators, as shown in Fig. \ref{fig:FLI}(a).  Each unit cell consists of three microrings, labeled A, B, C, with identical resonance frequencies. We define a unit cell as shown by the dashed lines in Fig. \ref{fig:FLI}(a).  This choice of unit cell leads to simple Floquet modes contained in a single cell in the perfect coupling limit.
Each microring is evanescently coupled to its neighbors with identical coupling strength denoted by a coupling angle $\theta$, which corresponds to power coupling coefficient $\kappa^2=\sin^2 \theta$.  We neglect back scattering so that each microring supports only a single mode of propagation, either clockwise or counterclockwise.
As light circulates around each ring, it couples periodically to neighbor resonators so that the FLI can be regarded as a periodically-driven system of dimension (2 + 1)D. Here the coordinate $z$ along each microring waveguide takes the role of time and the driving period is equal to the microring circumference $L$.  Over each period, the propagation of light can be decomposed into 4 coupling steps, as shown in Fig. 1(b). 
Using waveguide coupled mode theory, we can describe the evolution of light in each unit cell in terms of a Schrodinger-like equation \cite{song2024}
\begin{equation}
    -i\frac{\partial}{\partial z}|\psi(\textbf{k},z)\rangle=[\beta I + H_{\mathrm{FB}}(\textbf{k},z)]|\psi(\textbf{k},z)\rangle 
    \label{eq:Schrod_eq}
\end{equation}
where $\beta$ is the propagation constant of the microring waveguides, $I$ is the $3\times 3$ identity matrix, $\textbf{k} = (k_x, k_y)$ is the crystal momentum, and $|\psi(\textbf{k},z)\rangle = [\psi_{\mathrm{A}}, \psi_{\mathrm{B}}, \psi_{\mathrm{C}} ]^{\mathrm{T}}$ is the wave function representing the fields in the three microrings in a unit cell. The Floquet-Bloch Hamiltonian $H_{\mathrm{FB}}$ describes a periodic sequence of four coupling steps, $H_{\mathrm{FB}}(\textbf{k},z)=H_1\rightarrow H_2\rightarrow H_3\rightarrow H_4$.  The Hamiltonian $H_j$ in step $j$ is independent of $z$ and is given by
\begin{align}
    H_j(\textbf{k}) & =\begin{bmatrix}
        0& k_ce^{i k_x}e^{i s_j k_y} & 0\\
        k_c e^{-i k_x} e^{-i s_j k_y} &0 & 0\\
        0 & 0 & 0 \\
    \end{bmatrix},  j = 1, 3 \nonumber \\
     H_j(\textbf{k}) & =\begin{bmatrix}
        0 & 0 & 0\\     
        0 & 0 & k_c e^{-i s_j k_x}e^{-i k_y}\\
        0 & k_ce^{i s_j k_x}e^{i k_y}&0 \\
    \end{bmatrix},  j = 2, 4  
\end{align}
where $k_c = 4\theta / L$ is the coupling strength per unit length, $s_j = 0$ for $j = 1, 4$ and $s_j = 1$ for $j = 2, 3$.  The evolution operator of the system is given by $\mathcal{U}(\textbf{k},z) = e^{i \beta z} \mathcal{P} e^{i\int_0^z H_{FB}(\textbf{k},z')dz'}$, where $\mathcal{P}$ is the path-ordering operator which accounts for the order of the coupling steps.
Since the Hamiltonian $H_{\mathrm{FB}}$ is constant in each step, we can define the evolution operator for coupling step $j$ as $\mathcal{U}_j(\textbf{k},z)=e^{iH_j (z - z_j)}$, where $z_j = z - (j - 1)L/4$ marks the start of step $j$. 
Over one period, the system evolution is captured stroboscopically by the Floquet operator $U_{\mathrm{F}}(\textbf{k})=e^{i \beta L}U_{\mathrm{C}}(\textbf{k})$, where $U_{\mathrm{C}}$ is the effective coupling matrix,
\begin{equation}  
    U_{\mathrm{C}}(\textbf{k})=\mathcal{U}_{4}(\textbf{k},L/4) \mathcal{U}_{3}(\textbf{k},L/4) \mathcal{U}_{2}(\textbf{k},L/4) \mathcal{U}_1 (\textbf{k},L/4).
    \label{eq:coupling_matrix}
\end{equation}
Since a field $|\psi(\textbf{k},z)\rangle$ must return to itself after completing each roundtrip in the microrings, it satisfies the periodic boundary condition $|\psi(\textbf{k},z+L)\rangle = U_{\mathrm{F}}(\textbf{k})|\psi(\textbf{k},z)\rangle = e^{2im\pi}|\psi(\textbf{k},z)\rangle$, $m \in \mathcal{Z}$.  From this condition we obtain the eigenvalue equation for the Floquet-Bloch (FB) modes $|\Phi_n(\textbf{k})\rangle$,
\begin{equation}  
    U_{\mathrm{C}}(\textbf{k})|\Phi_n(\textbf{k})\rangle=e^{i\varepsilon_n(\textbf{k}) L}|\Phi_n(\textbf{k})\rangle
\end{equation}
where $\varepsilon_n = 2m\pi/L - \beta_n(\textbf{k})$ is the $n^{\mathrm{th}}$ quasienergy band.
Over the first Floquet-Brillouin zone, $\varepsilon \in [-\pi/L, \pi/L]$, which corresponds to one microring's free spectral range (FSR), the FLI has a flat band at zero quasienergy ($\varepsilon_0 = 0$), and two symmetric dispersive bulk bands $\varepsilon_{\pm}(\textbf{k})$, as shown in Fig. 1(c).
Figure 1(d) shows the quasienergy spectrum, projected over all $k_x, k_y$ values, vs. the coupling angle. The Chern numbers of the bands and winding numbers of the gaps are also shown on the plot.  At the critical coupling angle $\theta_c \sim 0.36 \pi$ \cite{song2024}, the two dispersive bands touch at the $\Gamma$  point ($\textbf{k} = (0, 0)$), which also marks a topological phase transition where the winding number of the $\pi$-gap changes abruptly from 0 for $\theta < \theta_c$ to 1 for $\theta > \theta_c$, while the Chern numbers of the dispersive bands change from $\pm 1$ to 0.  The topological phase in the region $\theta > \theta_c$ is called anomalous Floquet insulator, where the lattice has nontrivial winding numbers in the gaps even though all the bands have zero Chern number.
When $\theta = \pi/2$, which corresponds to $100 \%$ power coupling between neighbor rings, the dispersive bands become flattened so that the lattice has all flat bands at quasienergies $\varepsilon_n = 0, \pm2\pi/3 L$ and all the bulk states are localized.  This ABF lattice is of particular interest since analytical expressions for the Floquet states and associated topological invariants can be derived, which are given in Section I of the Supplemental Material (S.M. I).

\section{Displacements of the Floquet modes}

We investigate the micro-motions of the Floquet modes in an FLI lattice by tracking the displacements of their Wannier centers in real space over each evolution period.  Suppose the system starts out in an initial state $|\psi(\textbf{k},0)\rangle$ at $z=0$, which can be expressed in the basis of the Floquet-Bloch modes as $|\psi(\textbf{k},0)\rangle=\sum_n c_n |\Phi_{n}(\textbf{k})\rangle$.
The evolution of the state over each period is given by 
\begin{equation}
    |\psi(\textbf{k},z)\rangle = \mathcal{U}(\textbf{k},z)|\psi(\textbf{k},0)\rangle = \sum_{n=1}^3 c_n |\Psi_{n}(\textbf{k},z)\rangle
    \label{eq:ep}
\end{equation}
where $|\Psi_{n}(\textbf{k},z)\rangle = \mathcal{U}(\textbf{k},z)|\Phi_{n}(\textbf{k})\rangle$ are the $z$-evolved Floquet modes.  Using $\beta_n(\textbf{k}) = 2m\pi/L - \varepsilon_n(\textbf{k})$, we can express the evolution operator $\mathcal{U}$ as
\begin{align}
    \mathcal{U}(\textbf{k},z)  & = \Big \{ \sum_n e^{i\beta_nz} |\Phi_n\rangle \langle\Phi_n| 
    \Big \} e^{i\int_0^z H_{FB}(\textbf{k},z')dz'} \notag \\ 
    & = \sum_n e^{i(2m\pi/L - \varepsilon_n)z} |\Phi_n\rangle \langle\Phi_n| \mathcal{U}_C(\textbf{k},z)
\end{align}
where $\mathcal{U}_C = e^{i\int_0^z H_{FB}dz'}$ is the coupling matrix. It can be verified that the evolution operator is periodic, $\mathcal{U}(\textbf{k},L) = \mathcal{U}(\textbf{k},0) = I$. The $z$-evolved Floquet modes can then be computed from
\begin{equation}
    |\Psi_{n}(\textbf{k},z)\rangle = e^{2im\pi z/L}e^{-iH_{\mathrm{eff}}(\textbf{k})z} \mathcal{U}_C(\textbf{k},z)|\Phi_{n}(\textbf{k})\rangle
    \label{eq:Floquet_mode}
\end{equation}
where $H_{\mathrm{eff}} = \sum_n \varepsilon_n |\Phi_n\rangle\langle\Phi_n|$  is the effective Hamiltonian of the FLI. 

The displacement of the Wannier center $\textbf{r}_{\mathrm{c}} = [x_{\mathrm{c}}, y_{\mathrm{c}}]^{\mathrm{T}}$ of a state $|\psi\rangle$ is given by \cite{Nakagawa2020}
\begin{equation}
    \textbf{r}_{\mathrm{c}}(z)=\frac{1}{4\pi^2} \int_{BZ}\langle \psi(\textbf{k},z)|i\nabla_{\textbf{k}}| \psi(\textbf{k},z)\rangle d^2\textbf{k} 
    \label{eq:Wannier_center}
\end{equation}
Using the Floquet mode expansion in Eq.(\ref{eq:ep}), we obtain the instantaneous position of the Wannier center at a point $z$ in the evolution cycle as
\begin{equation}
    \textbf{r}_{\mathrm{c}}(z) =\frac{1}{4\pi^2}\int_{BZ} \textbf{c}^\dagger \textbf{A}_{\textbf{k}} \textbf{c} d^2 \textbf{k} = \sum_{m = 1}^3 \sum_{n = 1}^3 \frac{1}{4\pi^2}\int_{BZ} c_m^\dagger c_n [\textbf{A}_{\textbf{k}}]_{mn} d^2\textbf{k} \label{eq:rc}
\end{equation}
where $\textbf{c}=[c_1,c_2,c_3]^{\mathrm{T}}$ and $\textbf{A}_{\textbf{k}} = (A_{k_x}, A_{k_y})$ are the non-Abelian connection matrices with elements \cite{Anandan1988}
\begin{equation}
    [A_{\nu}]_{mn}=i\langle\Psi_m (\textbf{k},z)|\partial_{\nu}\Psi_n (\textbf{k},z)\rangle, \nu \in \{k_x, k_y\}
\end{equation}
Equation (\ref{eq:rc}) gives the instantaneous non-Abelian displacement of the Wannier center due to a non-adiabatic drive.  

If the system starts out in a Floquet-Bloch mode, $|\psi(\textbf{k},0)\rangle = |\Phi_n(\textbf{k})\rangle$, then $\textbf{c}$ is just the standard basis vector with $c_n = 1$.  In this case, the instantaneous displacement of the Wannier center of the $z$-evolved Floquet mode $|\Psi_n\rangle$ is obtained just from the Berry connections $[A_{k_x}]_{nn}$ and $[A_{k_y}]_{nn}$, 
\begin{equation}
    \textbf{r}_{\mathrm{c},n}(z)=\frac{1}{4\pi^2} \int_{BZ} [\textbf{A}_{\textbf{k}}]_{nn}d^2\textbf{k}
\end{equation}
Using $|\Psi_n\rangle$ in Eq.(\ref{eq:Floquet_mode}) to calculate the Berry connections, we obtain the trajectory of the center of Floquet mode $n$, which can be expressed as $\textbf{r}_{\mathrm{c},n}(z) = \textbf{r}_{\mathrm{c},n}(0) + \Delta\textbf{r}_{\mathrm{g},n}(z) + \Delta\textbf{r}_{\mathrm{d},n}(z)$, where 
\begin{align*}
    \textbf{r}_{\mathrm{c},n}(0) & = \frac{i}{4\pi^2} \int_{BZ} \langle\Phi_n|\grad_{\textbf{k}}|\Phi_n\rangle d^2\textbf{k} \notag \\
    \Delta\textbf{r}_{\mathrm{g},n}(z) & = \frac{i}{4\pi^2} \int_{BZ} \langle\Phi_n| \mathcal{U}_C^\dagger \grad_{\textbf{k}} \mathcal{U}_C |\Phi_n\rangle d^2\textbf{k} \notag \\
    \Delta\textbf{r}_{\mathrm{d},n}(z) & = \frac{i}{4\pi^2} \int_{BZ} \langle\Phi_n| \mathcal{U}_C^\dagger (e^{i H_{\mathrm{eff}} z} \grad_{\textbf{k}} e^{- i H_{\mathrm{eff}} z}) \mathcal{U}_C |\Phi_n\rangle d^2\textbf{k}
\end{align*}
In the above expressions, $\textbf{r}_{\mathrm{c},n}(0)$ is the initial displacement while $\Delta\textbf{r}_{\mathrm{g},n}(z)$ and $\Delta\textbf{r}_{\mathrm{d},n}(z)$ can be regarded as the displacements arising from the geometric and dynamic phases, respectively.   

\begin{figure}[t]
    \centering\includegraphics[scale=0.38]{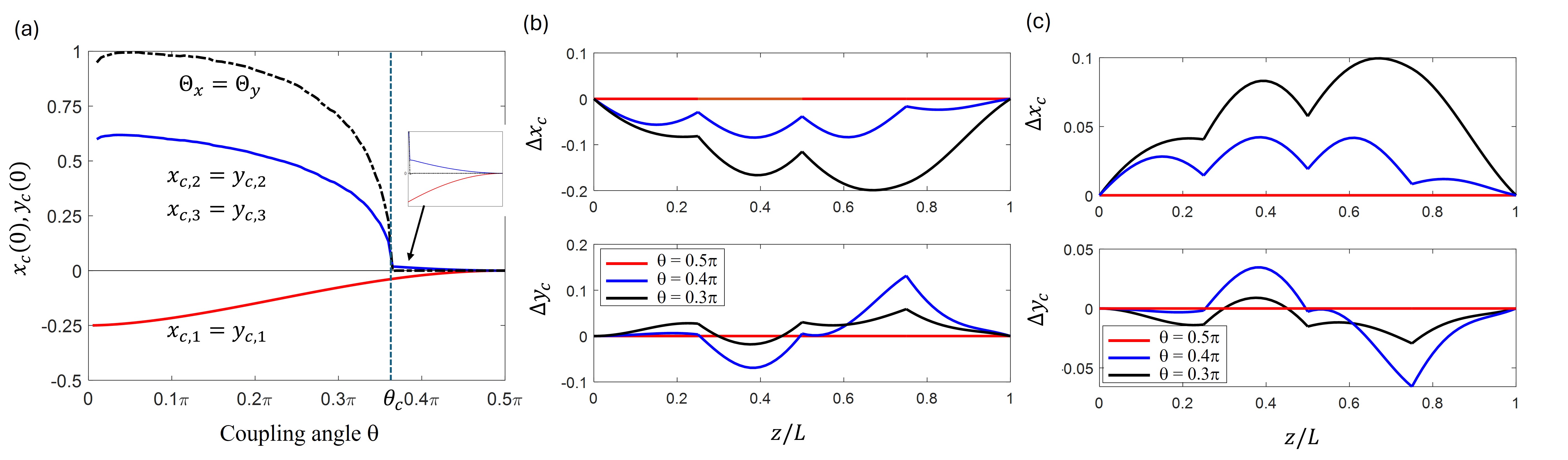}
 \caption{(a) Initial displacements $x_{\mathrm{c},n}(0) = y_{\mathrm{c},n}(0)$ of the Wannier center of Floquet mode $|\Phi_n\rangle$ vs. the coupling angle $\theta$. The sums of the displacements in $x$ and $y$ are indicated by the 2D Zak phases $\Theta_x$ and $\Theta_y$, respectively.  (b) and (c) Trajectories $\Delta x_{\mathrm{c},n}(z)$ (upper panel) and $\Delta y_{\mathrm{c},n}(z)$ (lower panel) of the Wannier center vs. $z$ over one period for (b) flat-band mode ($|\Psi_1\rangle$) and (c) dispersive modes ($|\Psi_2\rangle$ and $|\Psi_3\rangle$) for various coupling angles.}
\label{fig:wannier_center}
\end{figure}

Figure \ref{fig:wannier_center}(a) shows the initial displacements $x_{\mathrm{c},n}(0), y_{\mathrm{c},n}(0)$ of the Floquet-Bloch mode $|\Phi_n\rangle$ as functions of the coupling angle $\theta$.  The $x$ and $y$ displacements are equal for all three modes $(x_{\mathrm{c},n} = y_{\mathrm{c},n})$, with the displacements of the two dispersive modes $|\Phi_2\rangle$ and $|\Phi_3\rangle$ also identical $(\textbf{r}_{\mathrm{c},2} = \textbf{r}_{\mathrm{c},3})$. 
For $\theta > \theta_c$, the Wannier centers of the modes remain localized in the home unit cell around (0,0) but diverge outward as the coupling angle decreases below $\theta_c$, signifying distinct behaviors across the topological phase transition point at $\theta_c$.  The 2D Zak phases $\Theta_{x}$ and $\Theta_{y}$ \cite{Liu2017}, which are equal to the sums of the displacements in $x$ and $y$, respectively, are also seen to become invariant for $\theta > \theta_c$ $(\Theta_x = \sum_n x_{c,n}(0) = 0, \Theta_y = \sum_n y_{c,n}(0) = 0)$.  
As the Floquet mode $|\Psi_n\rangle$ evolves along $z$, its Wannier center is displaced from the initial position by $\Delta \textbf{r}_{\mathrm{c},n}(z) = \textbf{r}_{\mathrm{c},n}(z) - \textbf{r}_{\mathrm{c},n}(0)$.  The trajectories $\Delta \textbf{r}_{\mathrm{c},n}(z)$ of the Wannier centers of the flat band mode and dispersive modes are shown in Figs. \ref{fig:wannier_center}(b) and (c), respectively, for several coupling angles.  For the ABF lattice ($\theta = 0.5\pi$), the Wannier centers of all three modes exhibit zero displacement over each evolution period, which confirms that the three localized flat-band modes remain stationary over the entire cycle.  However, when $\theta < 0.5\pi$, even the flat-band mode at zero-quasienergy ($|\Psi_1\rangle$) exhibits motion in $x$ and $y$ during the evolution period, although it returns to its initial position after each cycle. This is because the flat quasienergy spectrum only captures the effective motion of the mode over each cycle and does not reveal its micro-motion during the evolution period.  This behavior is distinct from flat-band modes in static systems, whose Wannier centers always remain stationary.  The non-zero displacement of the flat-band mode in the periodically-driven system also implies that mixing with dispersive bands occurs during the evolution period.

For the flat-band modes in the ABF lattice, even though their Wannier centers do not execute any motion in $x$ or $y$, these modes do exhibit helical twists about their centers as they evolve along $z$, which can only be captured by taking into account the non-Abelian nature of the displacement operator.  In the next section, we show that this helical motion is characterized by a quantized non-Abelian helicity.

\section{Quantized non-Abelian helicity of the Floquet modes}

The helicity of a magnetic field enclosed in a volume $V$ is defined as $\mathcal{H} = \int \mathcal{A} \cdot \grad \cross \mathcal{A} dV$, where $\mathcal{A}(\textbf{r})$ is the magnetic vector potential \cite{berger1984}.  Here we generalize the helicity to the case where the components of the potential are matrix-valued,
\begin{equation} 
    \mathcal{H} = \frac{1}{4\pi^2} \int_0^L dz \int_{BZ} \mathrm{Tr}\{\textbf{A} \cdot \grad_{\textbf{k}} \cross \textbf{A} \} d^2\textbf{k} 
    \label{eq:nonA_helicity}
\end{equation}
where, for the FLI lattice, $\textbf{A}(\textbf{k},z)$ has components $(A_{k_x},A_{k_y},A_z)$.  The connection matrices $A_{\nu}$, $\nu \in \{k_x, k_y, z\}$, have elements $[A_{\nu}]_{m,n} = i\langle\Psi_m | \partial_{\nu} \Psi_n\rangle$.  
In component form, Eq.(\ref{eq:nonA_helicity}) can also be expressed as (see S.M. II)
\begin{align}
    \mathcal{H} &= -\frac{3}{4\pi^2}\int_0^L dz \int_{BZ} \text{Tr}\{A_{k_x}\partial_z A_{k_y} - A_{k_y}\partial_z A_{k_x}\} d^2\textbf{k}
    \label{eq:helicity}
\end{align}
By identifying $A_{k_x}, A_{k_y}$ with the position operators, $A_{k_x} \leftrightarrow \Tilde{\textbf{x}}$ and $A_{k_y} \leftrightarrow \Tilde{\textbf{y}}$, we can express the helicity as
\begin{equation}
    \mathcal{H} = -\frac{3}{4\pi^2}\int_0^L dz \int_{BZ} \text{Tr}\{\Tilde{\textbf{r}}\times \partial_z \Tilde{\textbf{r}}\cdot \hat{\textbf{z}}\} d^2\textbf{k}
    \label{eq:helicity2}
\end{equation}
The operator $\Tilde{\textbf{r}}\times \partial_z \Tilde{\textbf{r}}$ describes the twisting of the field lines.  
A similar operator (with $z$ replaced by time $t$) has also been used to calculate the orbital magnetization in 2D Floquet insulators \cite{nathan2017quantized}.

\begin{figure}[t]
\centering\includegraphics[scale=0.6]{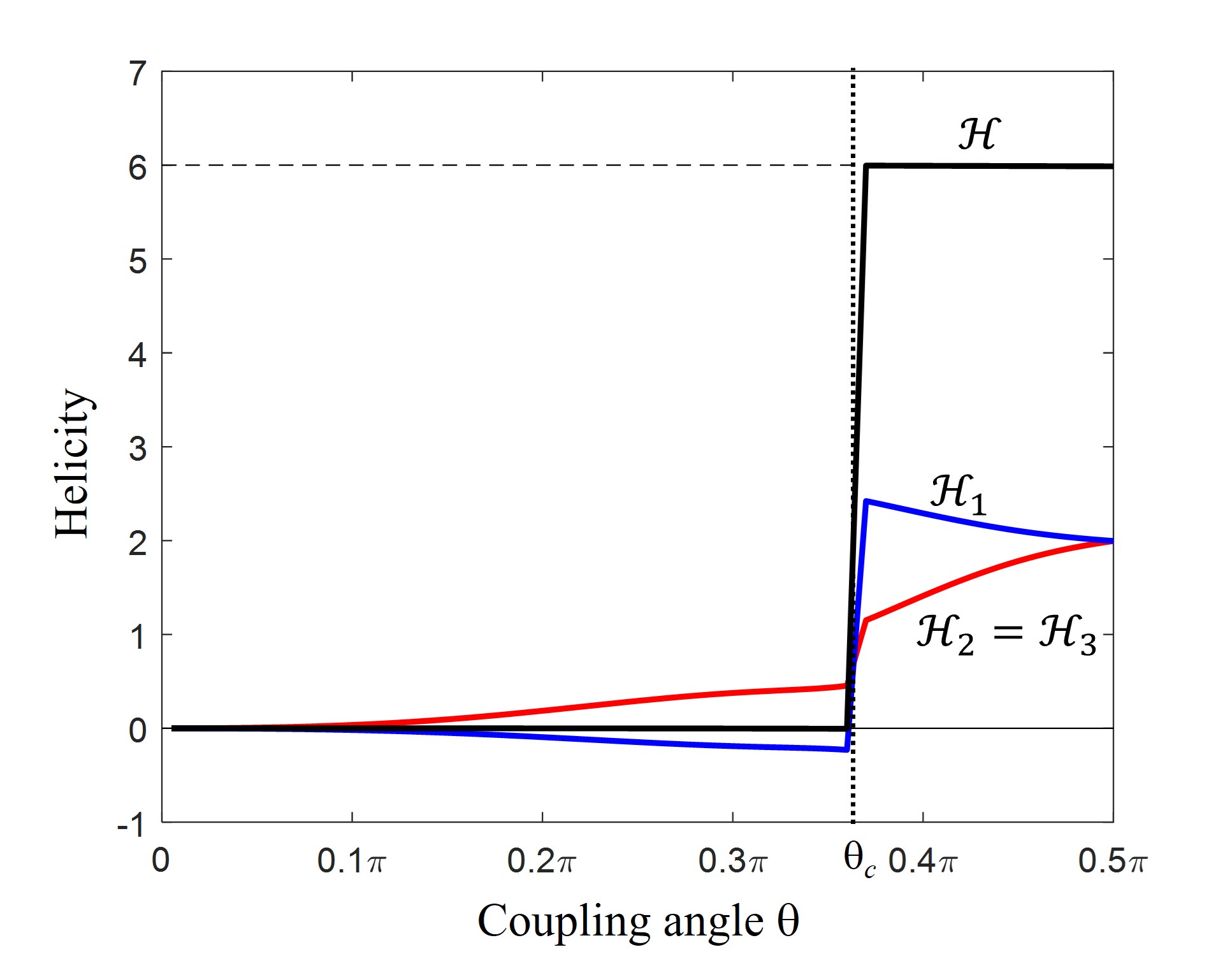}
 \caption{Helicity $\mathcal{H}$ of the Floquet modes of an FLI lattice as function of the coupling angle.  The helicity contributions of the flat-band mode $(\mathcal{H}_1)$ and the two dispersive modes $(\mathcal{H}_2 = \mathcal{H}_3)$ are also shown.}
    \label{fig:helicity}
\end{figure}

Using the $z$-evolved Floquet modes in Eq. (\ref{eq:Floquet_mode}), we computed the connection matrices $A_{k_x}(\textbf{k},z)$ and $A_{k_y}(\textbf{k},z)$, which are then used in Eq.(\ref{eq:helicity}) to obtain the helicity.  Figure \ref{fig:helicity} shows the computed helicity of the FLI as a function of the coupling angle.  The helicity contribution $\mathcal{H}_n$ of each mode $|\Psi_n\rangle$ to the total helicity, $\mathcal{H} = \sum_n \mathcal{H}_n$, is also shown.  
While the helicity $\mathcal{H}_n$ of each mode is not discrete, the total helicity $\mathcal{H}$ is quantized, exhibiting a discrete jump from zero for $\theta < \theta_c$ to 6 for $\theta > \theta_c$.
For the ABF lattice, the helicity can be analytically computed to give $\mathcal{H}_n = 2$ for each flat-band mode (see S.M. I), yielding a total quantized helicity of $\mathcal{H} = 6$.  Thus, even though the Wannier center of each flat-band mode is stationary, it exhibits a helical twisting motion over each evolution period as indicated by the nontrivial helicity.

The nontrivial helicity arises from the non-Abelian displacements of the system, which account for the intermixing of the three bands during each evolution period.  
The quantized helicity is also topological in origin and can be related to the 3rd Chern-Simons (CS) form of a (2 + 1)D system \cite{Ryu2010},
\begin{equation}
    CS_3 = \frac{1}{8\pi^2}\int_0^L dz \int_{BZ} \text{Tr}\{\textbf{A} \cdot \grad_{\textbf{k}} \cross \textbf{A}  - \frac{2}{3} i\textbf{A} \cdot \textbf{A} \cross \textbf{A}\} d^2\textbf{k}
    \label{eq:CS3_form}
\end{equation}
In the Supplemental Material (S.M. II), we show that $\textbf{A} \cdot \grad_{\textbf{k}} \cross \textbf{A} = i \textbf{A} \cdot \textbf{A} \cross \textbf{A}$, so that Eq.(\ref{eq:CS3_form}) can be expressed as 
\begin{equation}
    CS_3 = \frac{1}{24\pi^2}\int_0^L dz \int_{BZ} \text{Tr}\{\textbf{A} \cdot \grad_{\textbf{k}} \cross \textbf{A}\} d^2\textbf{k} = \frac{1}{6}\mathcal{H}
\end{equation}
Thus the helicity is 6 times the third Chern number, which is also equal to the winding number of the Floquet system \cite{leng2022n}.  The correspondence of the helicity to the CS form implies that the helicity is also gauge invariant.

\section{FLI lattice in a synthetic magnetic field}

The quantized helicity of the Floquet modes manifests itself in the behavior of the FLI lattice subject to a synthetic magnetic field, which allows for the possibility of experimentally detecting the nontrivial helicity.  It was shown in \cite{nathan2017quantized} that in a 2D Floquet system in which all the bulk modes are localized by disorder, the time-averaged orbital magnetization density is quantized and equal to the winding number of the system.  Here we demonstrate a similar relationship between the quasienergy shifts induced by a synthetic magnetic field and the quantized helicity of a FLI lattice, and propose a scheme for measuring the helicity.  

When a FLI lattice is subject to a uniform perpendicular magnetic field, $\textbf{B} = B \hat{\textbf{z}}$, the off-diagonal elements of the Hamiltonian in each step acquire a Peierls phase, $H_{ab} \rightarrow H_{ab}\exp{i\int_{\textbf{r}_a}^{\textbf{r}_b} \mathcal{A} \cdot d\textbf{r}}$, where $\textbf{r}_{(a,b)}$ denotes the location of site $(a,b)$ and $\mathcal{A}$ is the magnetic vector potential. In the symmetric gauge $\mathcal{A} = \frac{1}{2}B \textbf{r} \cross \hat{\textbf{z}}$, the change in the Hamiltonian with respect to the magnetic field $B$ can be expressed as \cite{nathan2017quantized}
\begin{equation}
    \frac{\partial H}{\partial B} = -\frac{1}{2} (\Tilde{\textbf{r}}\cross \partial_z \Tilde{\textbf{r}}) \cdot \hat{\textbf{z}}
\end{equation}
Substituting the above result into Eq.(\ref{eq:helicity2}) we obtain for the helicity
\begin{equation}
    \mathcal{H} = \frac{3}{2\pi^2}\int_0^L dz \int_{BZ} \text{Tr}\bigg\{\frac{\partial H_{\mathrm{FB}}}{\partial B} \bigg\} d^2\textbf{k}  
    \label{eq:helicity3}
\end{equation}
The perturbed Hamiltonian $H_{\mathrm{FB}}(B)$ leads to a shift in the quasienergies of the FLI lattice.  To evaluate this shift, we note that the determinant of the effective coupling matrix in Eq.(\ref{eq:coupling_matrix}), $U_{\mathrm{C}} = \mathcal{T}e^{i\int_0^L H_{\mathrm{FB}} dz} = e^{i H_{\mathrm{eff}} L}$, is given by
\begin{equation}
    \det(U_\mathrm{C}) = e^{i\int_0^L \mathrm{Tr}\{H_{\mathrm{FB}}\} dz} = e^{i \mathrm{Tr}\{H_{\mathrm{eff}}\} L}
    \label{eq:det_UF}
\end{equation}
Taking the derivative of $\det(U_{\mathrm{C}})$ with respect to the magnetic field strength, we obtain
\begin{equation}
    \frac{\partial }{\partial B} \det(U_{\mathrm{C}})= i \det(U_{\mathrm{C}}) \int_0^L  \mathrm{Tr} \bigg \{ \frac{\partial H_{\mathrm{FB}}}{\partial B} \bigg \} dz = i \det(U_{\mathrm{C}}) \mathrm{Tr} \bigg \{ \frac{\partial H_{\mathrm{eff}}}{\partial B} \bigg \} L
\end{equation}
which gives
\begin{equation}
    \int_0^L \mathrm{Tr} \bigg\{ \frac{\partial H_{\mathrm{FB}}}{\partial B} \bigg \} dz =  \mathrm{Tr} \bigg\{ \frac{\partial H_{\mathrm{eff}}}{\partial B} \bigg \} L =  \sum_n \frac{\partial \varepsilon_n L}{\partial B} 
    \label{eq:trace_H}
\end{equation}
 where we have made use of the relation $\mathrm{Tr}\{H_{\mathrm{eff}}\} = \sum_n \varepsilon_n$. The helicity in Eq.(\ref{eq:helicity3}) can thus be computed as
\begin{equation}
    \mathcal{H} = \frac{3}{2\pi^2} \int_{BZ} \sum_n \frac{\partial \varepsilon_nL}{\partial B} d^2\textbf{k}
    \simeq \frac{3}{2\pi^2B} \int_{BZ} \sum_n \big [ \varepsilon_n(B)L - \varepsilon_n(0)L \big] d^2\textbf{k} 
    \label{eq:helicity4}
\end{equation}
Since the magnetic field only affects the off-diagonal elements of the Hamiltonian, $H_{\mathrm{FB}}(B)$ remains traceless so that we can conclude from Eq.(\ref{eq:trace_H}) that $\sum_n \varepsilon_n$ is independent of momentum $\textbf{k}$.
 This allows us to evaluate the integral over a BZ in Eq.(\ref{eq:helicity4}) to get
 \begin{equation}
    \mathcal{H} \simeq \frac{6}{B} \bigg[\sum_n \varepsilon_n(B)L -  \sum_n \varepsilon_n(0)L \bigg]
\end{equation}
The above result shows that the helicity can be measured from the total change in the quasienergies of the Floquet modes due to an applied uniform magnetic field $B$. It should be noted that for a static system, $H_{\mathrm{eff}}(B) = H(B)$, which is always traceless, so that $\sum_n \partial \varepsilon_n / \partial B = 0$. Thus while the individual energy bands of a static system will shift in the presence of a magnetic field, the total energy change $\sum \varepsilon_n(B)$ is zero, implying that the helicity of the system is always trivial. 

\begin{figure}
\centering\includegraphics[scale=0.6]{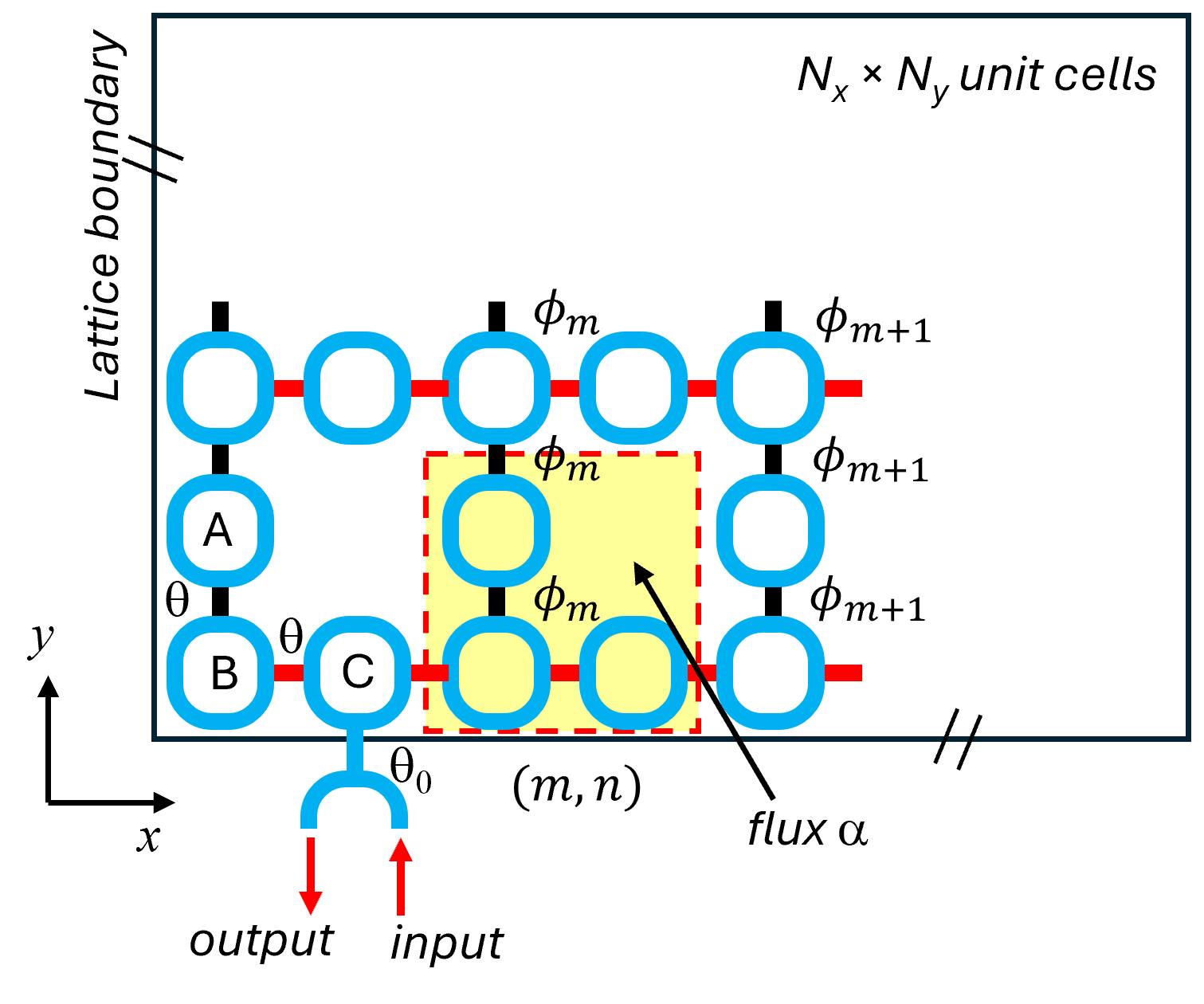}
 \caption{An FLI lattice with $N_x \times N_y$ square unit cells subject to a synthetic magnetic flux $\alpha$ generated by a coupling phase gradient $\phi_m = \pi\alpha m$ along the $x$ direction.}
\label{fig:FLI_in_mag_field}
\end{figure}

To demonstrate the behavior associated with a nontrivial helicity, we simulated a finite FLI lattice with $N_x \times N_y$ square unit cells in a magnetic field $B$ generated by the Landau gauge $\mathcal{A} = B x \hat{\textbf{y}}$.  
This gauge can be synthesized by introducing a coupling phase between rings A and B as shown in Fig. \ref{fig:FLI_in_mag_field}.  The coupling phase $\phi_m$ in cell $(m,n)$ is set to be $\phi_{m} = \pi \alpha m$, where $\alpha = Ba^2/2\pi$ is the flux threading a unit cell of area $a^2$. 
The coupling phase can be realized using link rings between site rings as in \cite{Hafezi2013}.    
To excite bulk modes in the FLI lattice, we couple light into ring C of a cell on the bottom lattice boundary via an input waveguide with coupling angle $\theta_0$.  The optical power transmitted at the output of the waveguide is computed using the method in \cite{afzal2018topological}.

\begin{figure}[t]
\centering\includegraphics[scale=0.5]{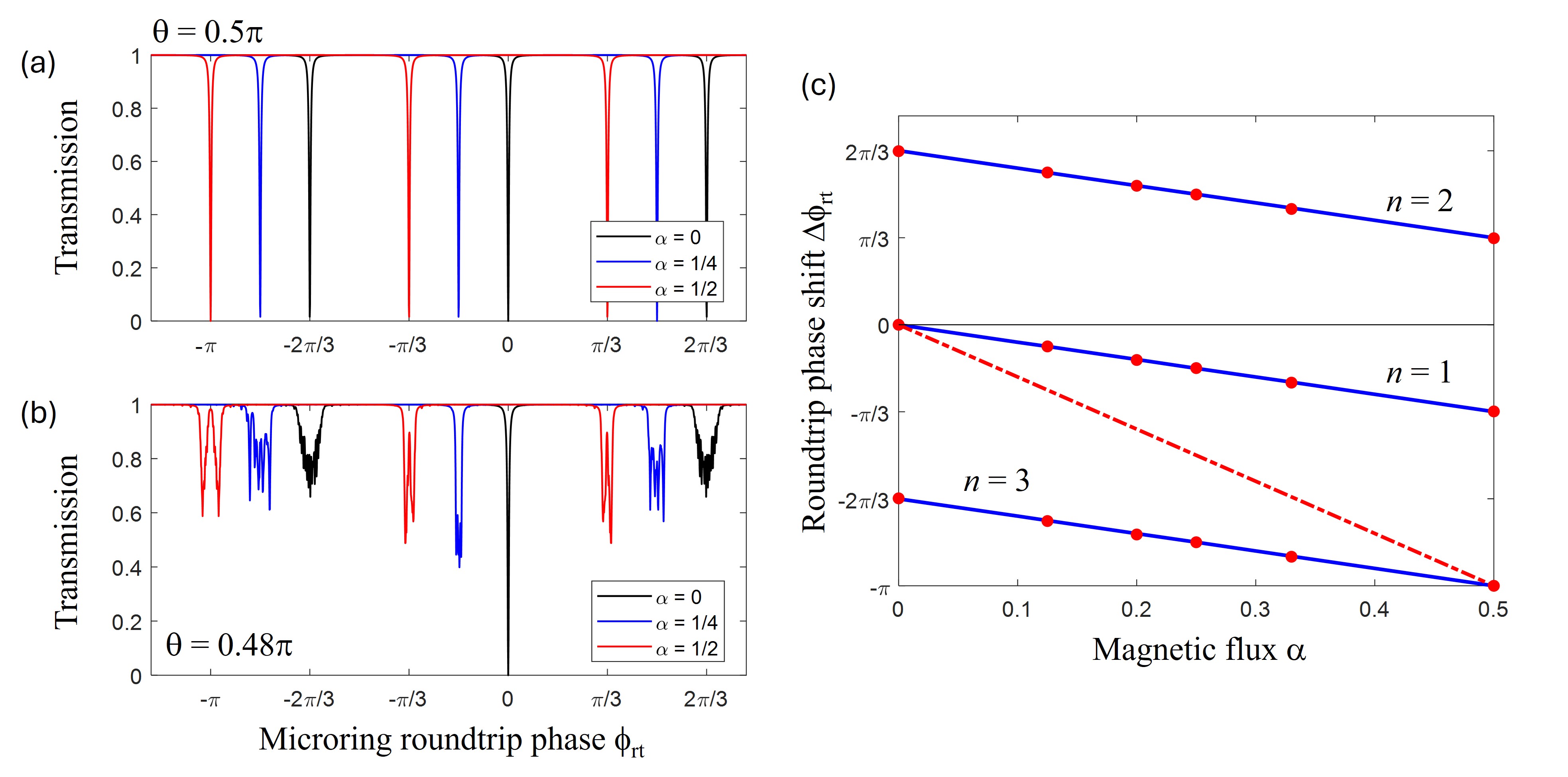}
 \caption{Simulated transmission spectra over one FSR of an FLI lattice subject to different values of applied magnetic flux $\alpha$ for (a) perfect coupling case ($\theta = 0.5\pi$) and (b) $\theta = 0.48\pi$. (c) Shifts in resonance frequencies (in terms microring roundtrip phase) of the Floquet modes $n$ as functions of the applied flux $\alpha$ for FLI lattice with $\theta = 0.5\pi$ (blue line) and $\theta = 0.48\pi$ (red dots).  The red dashed line is the sum of the roundtrip phase shifts of the three modes.}
\label{fig:helicity_meas}
\end{figure}

Figure \ref{fig:helicity_meas}(a) shows the transmission spectrum as function of the microring roundtrip phase of an ABF lattice with $N_x = 12, N_y = 12$ subject to  various flux values $\alpha$. The microring roundtrip phase can be related to the input light frequency via $\phi_{\mathrm{rt}} = n_{\mathrm{g}}(\omega/c)L$, where $n_{\mathrm{g}}$ is the group index of the microring waveguide.
The input waveguide coupling angle is set at $\theta_0 = 0.05\pi$ and the roundtrip field attenuation in each ring is assumed to be $a_{\mathrm{rt}} = \cos^{1/3}(\theta_0)$ to achieve critical coupling to the flat band resonance modes.  When there is no magnetic flux ($\alpha = 0$), the transmission spectrum (black trace) over one FSR consists of three sharp resonance dips corresponding to the three flat band modes at quasienergies $0, \pm 2\pi/3L$. When the flux is turned on, the three resonances are shifted by the same amount, as shown by the red and blue traces in Fig. \ref{fig:helicity_meas}(a).  We plot the resonance frequency shifts in terms the microring roundtrip phase as functions of the applied flux $\alpha$ in Fig. \ref{fig:helicity_meas}(c) (blue lines), which shows a linear dependence with a slope of $d\Delta\phi_{\mathrm{rt}}/d \alpha = -0.666\pi$.  
Since $\Delta \phi_{\mathrm{rt}} = - \varepsilon_n L$ and $\alpha = B/2\pi$ (assuming lattice constant $a = 1$), we obtain $d \varepsilon_n L / d B = -(1/2\pi) d\Delta\phi_{\mathrm{rt}}/d \alpha = 0.333$.  The total quasienergy change of the three modes is $\sum_n d \varepsilon_n L / d B = 1$, yielding a total helicity of $\mathcal{H} = 6$ in agreement with the value calculated from the non-Abelian connection matrices.

Figure \ref{fig:helicity_meas}(b) shows the transmission spectrum for the lattice with coupling angle slightly deviated from perfect coupling, $\theta = 0.48\pi$.  When there is no magnetic flux, the transmission spectrum (black trace) also shows three dips, but the dips at $\pm 2\pi/3L$ are broadened since the bands become slightly dispersive.  In the presence of a magnetic flux, the resonance dips exhibit the same shifts as the perfect coupling case, but each resonance also splits into multiple dips (in general there are $M$ dips for an applied flux $\alpha = 1/M$).  Nevertheless, we can determine the average shift for each band, which is shown by the red dots in Fig. \ref{fig:helicity_meas}(c). The quasienergy shifts have the same dependence on the flux as in the ABF lattice, yielding the same quantized helicity $\mathcal{H} = 6$.

\section{Braid structure of Floquet modes in an ABF lattice}

We showed in Section IV  that the flat-band modes of the ABF lattice have nontrivial quantized helicity even though their Wannier centers exhibit zero displacement over each evolution period.  The nontrivial helicity suggests that the modes undergo a twisting motion over each period that is not captured in the trajectory of the Wannier center.  Here we show that by tracking the micromotions of the individual basis components of a flat-band mode in real space, its evolution can be represented by a three-component braid with nontrivial winding number.  This braiding structure helps provide an intuitive picture of the nontrivial helicity of the FLI lattice.

The Floquet-Bloch modes of an ABF lattice can be analytically derived as $|\Phi_n\rangle = 1/\sqrt{3} [ -1,  ie^{-i\phi_n}, e^{i\phi_n} ]^\mathrm{T}, \phi_n \in \{0, \pm 2\pi/3\}$ (see S.M. I), which indicates that each mode consists of fields localized in waveguides A, B, and C of a single unit cell.
We can express these modes in the site basis as
\begin{equation}
    |\Phi_n (\textbf{k})\rangle = \tfrac{1}{\sqrt{3}} \big (- |A_{\textbf{k}}\rangle + ie^{-i\phi_n} |B_{\textbf{k}}\rangle + e^{i\phi_n} |C_{\textbf{k}}\rangle \big ) = \sum_m \alpha_{n,m}|m_{\textbf{k}}\rangle
\end{equation}
where $|m_{\textbf{k}}\rangle \in \{|A_{\textbf{k}}\rangle, |B_{\textbf{k}}\rangle, |C_{\textbf{k}}\rangle \}$ represents a field in waveguide $m$ and $\alpha_{n,m} = \langle m_{\textbf{k}} | \Phi_n \rangle$.   Starting at $z = 0$, the mode $|\Phi_n\rangle$ evolves as $|\Psi_n(\textbf{k},z)\rangle = \mathcal{U}(\textbf{k},z) |\Phi_n\rangle = \sum_{m} \alpha_{n,m} \mathcal{U} |m_{\textbf{k}}\rangle$.  The evolution of the flat-band mode thus consists of three components, $|\psi_m(\textbf{k},z)\rangle = \alpha_{n,m} \mathcal{U} |m_{\textbf{k}}\rangle$, each evolving from site $m \in \{A, B, C\}$ of the initial state.  For example, the component evolving from site $A$ over the 4 coupling steps in one period is (see S.M. I)

\begin{equation}
    |\psi_A(\textbf{k},z)\rangle = \frac{1}{\sqrt{3}} 
    \begin{cases}
         -\tau_{z,1}|A_{\textbf{k}}\rangle - i\kappa_{z,1} e^{-ik_x}|B_{\textbf{k}}\rangle & 0 \leq z < L/4 \\
          -i\tau_{z,2}e^{-ik_x}|B_{\textbf{k}}\rangle + \kappa_{z,2} e^{ik_y}|C_{\textbf{k}}\rangle & 
    L/4 \leq z < L/2 \\
     e^{ik_y}|C_{\textbf{k}}\rangle & L/2 \leq z < 3L/4 \\
      \tau_{z,4}e^{ik_y}|C_{\textbf{k}}\rangle - i\kappa_{z,4}|B_{\textbf{k}}\rangle & 3L/4 \leq z < L \\
    \end{cases}
\end{equation}

\begin{figure}[t]
\centering\includegraphics[width=0.6\linewidth]{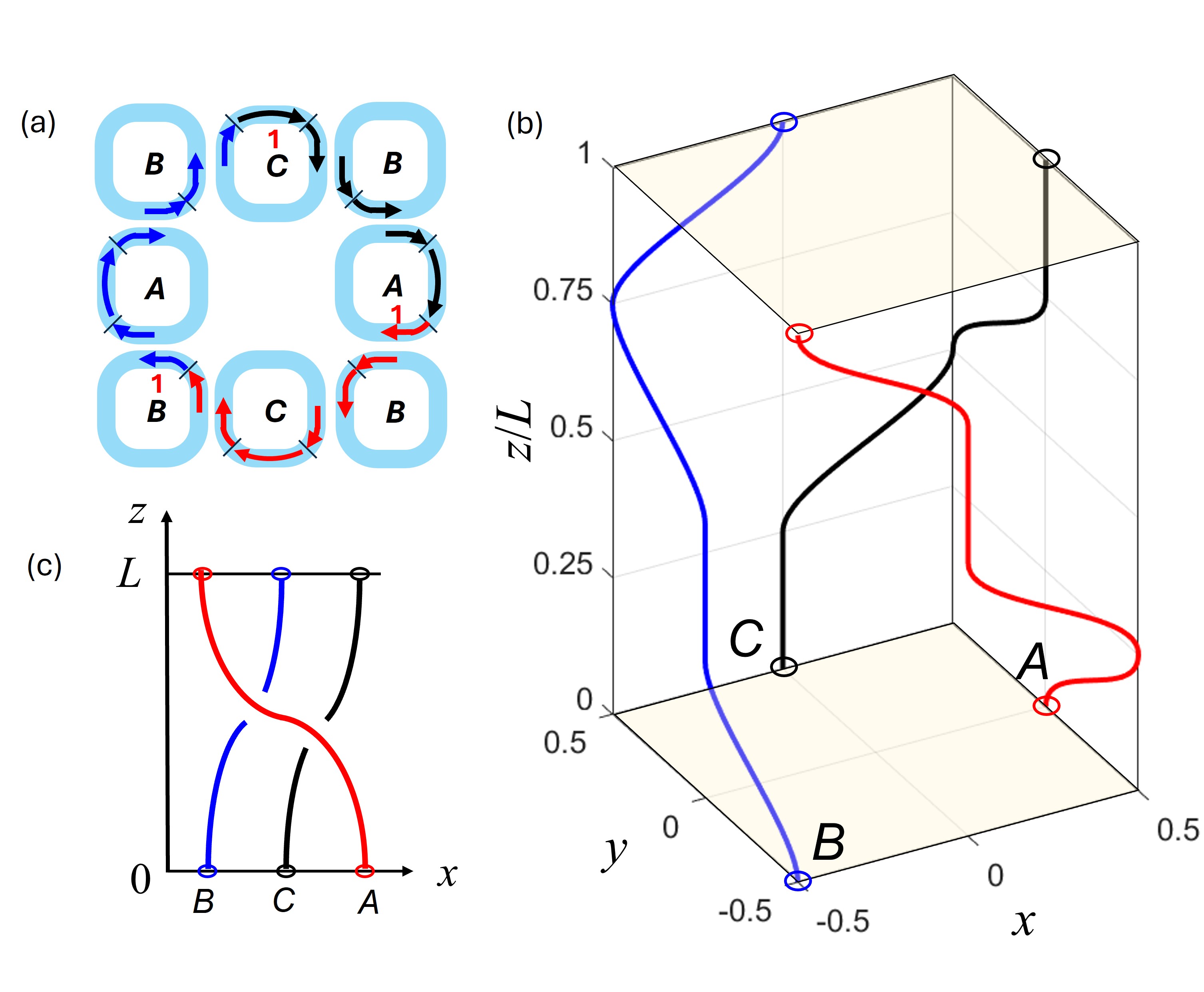}
 \caption{(a) Field pattern of a flat-band mode of the ABF lattice. (b) Braiding world lines showing the trajectories of the Wannier centers of the three components of the flat-band mode $|\Phi_n\rangle$ over one evolution period. (c) Three-component braid representation of the world lines projected onto the $x-z$ plane as viewed from the negative $y$ direction}
\label{fig:braid}
\end{figure}

The states evolving from sites $B$ and $C$ are computed in a similar way.
We can track the evolution of each component in real space by computing its Wannier function in the home unit cell $(0,0)$ (centered at $\textbf{R}_0 = [X_0, Y_0]^{\mathrm{T}} = [0, 0]^{\mathrm{T}}$) \cite{Nakagawa2020}
\begin{equation}
    |w_m(x, y, z)\rangle = \frac{1}{4\pi^2} \int_{BZ} e^{i\textbf{k}\cdot \textbf{r}} |\psi_m(\textbf{k},z)\rangle d^2 \textbf{k}
    \label{eq:Wannier_fcn}
\end{equation} 
For example, for component $|\psi_A(\textbf{k},z)\rangle$ above, the corresponding normalized Wannier function is

\begin{equation}
    |w_A(x,y,z)\rangle = \begin{cases}
         \tau_{z,1}|A_{0,0}\rangle + i\kappa_{z,1}|B_{1,0}\rangle & 0 \leq z < L/4 \\
          i\tau_{z,2}|B_{1,0}\rangle - \kappa_{z,2}|C_{0,-1}\rangle & L/4 \leq z < L/2 \\
            -|C_{0,-1}\rangle & L/2 \leq z < 3L/4 \\
               -\tau_{z,4}|C_{0,-1}\rangle + i\kappa_{z,4}|B_{0,0}\rangle & 3L/4 \leq z < L\\
    \end{cases}
\end{equation}

where $|M_{m,n}\rangle$ represents field in waveguide $M$ of unit cell $(m,n)$.  The Wannier functions of the other two components are given in the Supplemental Material (Section I). The complete field distribution of the flat-band mode in real space as mapped out by the Wannier functions $|w_m(x, y, z)\rangle$ over one period is shown in Fig. \ref{fig:braid}(a).
 The mode comprises of three strands which evolve to form a close loop over each period. This mode pattern has also been experimentally observed in \cite{Song2025}.
 The trajectory of each strand can be determined by computing the Wannier center of each component function $|w_m(x, y, z)\rangle$.  This is facilitated by defining the position operator $\tilde{\textbf{r}}$ such that $\langle M | \tilde{\textbf{r}} | M \rangle$ gives the position of site $M$.  Specifically, we have 
 \begin{align}
     \langle A_{m,n} | \tilde{\textbf{r}} | A_{m,n} \rangle & = \textbf{R}_{m,n} + [1/2, 0]^{\mathrm{T}} \notag \\
     \langle B_{m,n} | \tilde{\textbf{r}} | B_{m,n} \rangle & = \textbf{R}_{m,n} + [-1/2, -1/2]^{\mathrm{T}} \notag \\
     \langle C_{m,n} | \tilde{\textbf{r}} | C_{m,n} \rangle & = \textbf{R}_{m,n} + [0, 1/2]^{\mathrm{T}} \notag
 \end{align}
 where $\textbf{R}_{m,n} = [m, n]^{\mathrm{T}}$ is the center of cell $(m, n)$.  
The trajectories of the three components as given by the positions $\textbf{r}_{m,j}$ of their Wannier centers in each step $j$, $\textbf{r}_{m}(z) = \{\textbf{r}_{m,1}, \textbf{r}_{m,2}, \textbf{r}_{m,3}, \textbf{r}_{m,4}\}$, are (see S.M. I):
\begin{align}
    & \textbf{r}_{A}(z) = \left[\begin{array}{c}
        1/2\\
        -\kappa_{z,1}^2/2\\
    \end{array}\right] , 
    \left[\begin{array}{c}
        \tau_{z,2}^2/2\\
        -1/2\\
    \end{array}\right] ,
        \left[\begin{array}{c}
        0\\
        -1/2\\
    \end{array}\right] ,
    \left[\begin{array}{c}
        -\kappa_{z,4}^2/2\\
        -1/2\\
    \end{array}\right] \notag \\
    & \textbf{r}_{B}(z) =
    \left[\begin{array}{c}
        -1/2\\
        -\tau_{z,1}^2/2\\
    \end{array}\right] ,
    \left[\begin{array}{c}
        -1/2\\
        0\\
    \end{array}\right] ,
        \left[\begin{array}{c}
        -1/2\\
        \kappa_{z,3}^2/2\\
    \end{array}\right] ,
    \left[\begin{array}{c}
        -\tau_{z,4}^2/2\\
        1/2\\
    \end{array}\right] \notag \\
    & \textbf{r}_{C}(z) = 
    \left[\begin{array}{c}
        0\\
        1/2\\
    \end{array}\right]  ,
    \left[\begin{array}{c}
        \kappa_{z,2}^2/2\\
        1/2\
    \end{array}\right]  ,
    \left[\begin{array}{c}
        1/2\\
        \tau_{z,3}^2/2\\
    \end{array}\right] ,
    \left[\begin{array}{c}
        1/2\\
        0\\
    \end{array}\right]    
    \label{eq:Wannier_trajectories}
\end{align}
The trajectories $\textbf{r}_m(z)$ of the components are plotted in Fig. \ref{fig:braid}(b), which depicts three world lines twisting about each other, forming a three-component braid (Fig. \ref{fig:braid}(c)).  
 We can classify the braid by computing its winding number $W$ over each period \cite{berger1991third} .  By defining the world lines in terms of the complex-valued functions $\gamma_{\mu}(z) = x_{\mu}(z) + i y_{\mu}(z)$, we calculate the relative angles $\theta_{\mu,\nu}(z) = \mathrm{Im}\{\ln{(\gamma_{\mu} - \gamma_{\nu})}\}$ between strands $\mu, \nu \in \{A, B, C\}$.  
The pair-wise winding numbers are then found to be 
 \begin{align}
     W_{A,B} & = \theta_{A,B}(L) - \theta_{A,B}(0) = -\pi/2 - 2\tan^{-1}(1/2) \notag \\
     W_{B,C} & = \theta_{B,C}(L) - \theta_{B,C}(0) = -3\pi/4 + \tan^{-1}(1/2) = W_{C,A} \notag
 \end{align}
 The winding number of the three-component braid associated with mode $|\Phi_n\rangle$ is then $W_n = W_{A,B} + W_{B,C} + W_{C,A} = -2\pi$.  Summing over the three flat-band modes, we obtain $W = -6\pi$ for the total winding number of the ABF lattice. The helicity is related to the winding number by \cite{berger1991third}  $\mathcal{H} = -W/\pi = 6$, which is in agreement with the value computed from the non-Abelian connection matrices.

\section{Conclusion}

We showed in this work that flat bands in a 2D Floquet-Lieb topological photonic lattice can be characterized by a nontrivial quantized helicity even though they have zero Chern number.  This helicity arises from the helical micromotions of the modes, which are revealed through the trajectories of their non-Abelian displacements as a result of band mixing, even though all the quasienergy bands involved are non-degenerate.  We also showed that these trajectories can be represented by a braiding of their world lines, whose winding number is related to the helicity of the system. We proposed a scheme for measuring the non-Abelian helicity of an ABF microring lattice by subjecting it to a synthetic magnetic field and measuring the resonance frequency shifts of the flat-band modes.  As a future research direction, we can explore the possibility of embedding gratings into the rings to generate optical vortex beams emitted from the lattice whose topological properties can be directly measured.
Our work suggests that 2D Floquet topological insulators based on microring lattices can provide a versatile platform for investigating non-Abelian physics and strongly correlated phenomena associated with photonic flat-band systems.

\bibliography{references.bib}

\widetext
\clearpage
\begin{center}
\textbf{\large Quantized non-Abelian helicity of flat bands in 2D Floquet topological photonic insulators\\Supplemental Materials}
\end{center}
\setcounter{section}{0}
\renewcommand{\thefigure}{S\arabic{figure}}
\renewcommand{\theequation}{S\arabic{equation}}
\renewcommand{\thesection}{S-\Roman{section}}
\section{Helicity of all-bands-flat Floquet-Lieb microring lattice}
\subsection{Flat-band Floquet modes}

For the ABF lattice with perfect coupling ($\theta = \pi/2$), the evolution matrix for each coupling step $j$,  $\mathcal{U}_j(\textbf{k}, z) = e^{iH_j(\textbf{k})z}$, can be explicitly evaluated to give
\begin{align}
    \mathcal{U}_j(\textbf{k},z)=\begin{bmatrix}
        \tau_{z,j}& i\kappa_{z,j} e^{i k_x}e^{i s_j k_y} & 0\\
        i\kappa_{z,j} e^{-i k_x}e^{-i s_j k_y} &\tau_{z,j}&0\\
        0 & 0 & 1\\
    \end{bmatrix}, j = 1, 3 \nonumber \\
     \mathcal{U}_j(\textbf{k},z)=\begin{bmatrix}
        1 & 0 & 0 \\
        0 & \tau_{z,j} &i\kappa_{z,j} e^{-i s_j k_x}e^{-i k_y} \\
        0 & i\kappa_{z,j} e^{i s_j k_x}e^{i k_y}&\tau_{z,j} \\
    \end{bmatrix}, j = 2, 4 
    \label{eq:coupling_matrices}
\end{align}
where $\kappa_{z,j} = \sin[2\pi(z - z_j)/L], \tau_{z,j} = \cos[2\pi(z - z_j)/L]$, with $z_j = (j - 1)L/4$ marking the starting position of step $j$. The index $s_j$ is equal to 0 for steps $j = 1,4$ and 1 for steps $j = 2,3$. The Floquet operator is $U_\mathrm{F} = e^{i\beta L}U_\mathrm{C}$, where
\begin{align}
    U_\mathrm{C} & = \mathcal{U}_4(\textbf{k},L/4)\mathcal{U}_3(\textbf{k},L/4) \mathcal{U}_2(\textbf{k},L/4)\mathcal{U}_1(\textbf{k},L/4) \nonumber \\
    & = \begin{bmatrix}
        0 & 0 & -1 \\
        -i & 0 & 0 \\
        0 & -i & 0 
    \end{bmatrix} \nonumber
\end{align}
The eigenvalues of $U_\mathrm{F}$ are $e^{i (\beta_n L + \phi_n)} = e^{2im\pi}, \phi_n = \varepsilon_n L = \{0, \pm 2\pi/3\}$, with eigenfunctions (Floquet-Bloch modes)
\begin{equation}
    |\Phi_n\rangle = \frac{1}{\sqrt{3}} \begin{bmatrix} -1, & ie^{-i\phi_n}, & e^{i\phi_n} 
    \end{bmatrix}^\mathrm{T} \\
    \label{eq:flat_band_modes}
\end{equation}
Since $|\Phi_n\rangle$ is independent of $\textbf{k}$, the initial displacement of the Wannier center of the flat-band mode is zero:
\begin{equation}
    \textbf{r}_{\mathrm{c},n}(0) = \frac{i}{4\pi^2}\int_{BZ} \langle \Phi_n | \grad_{\textbf{k}} | \Phi_n \rangle d^2\textbf{k} = [0, 0]^{\mathrm{T}}
\end{equation}
The effective Hamiltonian of the ABF lattice is 
\begin{equation}
    H_{\mathrm{eff}} = \sum_{n = 1}^3 \varepsilon_n |\Phi_n\rangle \langle \Phi_n | = - \frac{2\pi}{3\sqrt{3} L}\begin{bmatrix}
        0 & 1 & -i \\
        1 & 0 & 1 \\
        i & 1 & 0 \\
    \end{bmatrix}
\end{equation}
which is also independent of $\textbf{k}$.

\subsection{Non-Abelian displacement of the flat-band modes}

The non-Abelian connection matrices $A_{k_x}(\textbf{k},z)$ and $A_{k_y}(\textbf{k},z)$ are computed using the $z$-evolved Floquet states $|\Psi_n(\textbf{k},z)\rangle$ in Eq. (7) in the main text.  Since $H_{\mathrm{eff}}$ is independent of $\textbf{k}$ for the ABF lattice, the $(m,n)$ element of the connection matrices is obtained just from the coupling matrix $\mathcal{U}_C(\textbf{k},z)$,
\begin{equation}
    [A_{\nu}]_{m,n}= i \langle \Phi_m | \mathcal{U}_C^{\dagger} \partial_{\nu} \mathcal{U}_C |\Phi_n \rangle
    \label{eq:connection_matrices}
\end{equation}
where $\nu \in \{k_x, k_y\}$.  It is convenient to define the non-Abelian displacement operators $\Delta \tilde{\textbf{x}} = i\mathcal{U}_C^{\dagger} \partial_{k_x} \mathcal{U}_C$ and $\Delta \tilde{\textbf{y}} = i\mathcal{U}_C^{\dagger} \partial_{k_y} \mathcal{U}_C$ so that $[A_{k_x}]_{m,n} = \langle \Phi_m | \Delta \tilde{\textbf{x}} |\Phi_n\rangle$ and $[A_{k_y}]_{m,n} = \langle \Phi_m | \Delta \tilde{\textbf{y}} |\Phi_n\rangle$.  The displacement operators in each step are computed as follows:

In step 1, we have $\mathcal{U}_C(\textbf{k},z) = \mathcal{U}_1(\textbf{k},z)$, where $\mathcal{U}_1$ is given in Eq.(\ref{eq:coupling_matrices}).  The displacement operators are thus (only the $z$ dependence is explicitly indicated)
\begin{align}
    \Delta \tilde{\textbf{x}}(z) & =  i \mathcal{U}_1^{\dagger} \partial_{k_x} \mathcal{U}_1 = \Delta \tilde{\textbf{x}}_1(z) \notag \\
    \Delta \tilde{\textbf{y}}(z) & = i \mathcal{U}_1^{\dagger} \partial_{k_y} \mathcal{U}_1 = 0 
\end{align}
with
\begin{equation}
    \Delta \tilde{\textbf{x}}_1(z) = \begin{bmatrix}
        \kappa_{z,1}^2 & -i\kappa_{z,1} \tau_{z,1} e^{i k_x} & 0 \\
        i \kappa_{z,1} \tau_{z,1} e^{-i k_x} & -\kappa_{z,1}^2 & 0 \\
        0 & 0 & 0 \\        
    \end{bmatrix}
    \label{eq:displacement_op_dx1}
\end{equation}
The displacement of the Wannier center of Floquet mode $|\Phi_n\rangle$ in step 1 is then
\begin{align}
    \Delta x_{\mathrm{c},n}(z) & = \frac{1}{4\pi^2}\int_{BZ} \langle\Phi_n | \Delta\tilde{\textbf{x}}_1(z) |\Phi_n \rangle d^2\textbf{k}\notag \\                  
    & = -\frac{1}{4\pi^2}\int_{BZ} \frac{2}{3}\kappa_{z,1}\tau_{z,1} \cos(k_x - \phi_n) d^2\textbf{k} = 0 
    \label{eq:displacement_step1}
\end{align}
and $\Delta y_{\mathrm{c},n}(z) = 0$.  

In step 2, we have $\mathcal{U}_C(\textbf{k},z) = \mathcal{U}_2(\textbf{k},z)\mathcal{U}_1(\textbf{k},L/4)$.  The displacement operators are
\begin{align}
    \Delta \tilde{\textbf{x}}(z) & =  i \mathcal{U}_1^{\dagger} \partial_{k_x} \mathcal{U}_1 + i \mathcal{U}_1^{\dagger}(\mathcal{U}_2^{\dagger} \partial_{k_x} \mathcal{U}_2) \mathcal{U}_1  = \Delta \tilde{\textbf{x}}_1(L/4) + \Delta\tilde{\textbf{x}}_2(z) \notag \\
    \Delta \tilde{\textbf{y}}(z) & = i\mathcal{U}_1^{\dagger}(\mathcal{U}_2^{\dagger} \partial_{k_y} \mathcal{U}_2) \mathcal{U}_1 
    = \Delta\tilde{\textbf{y}}_2(z) \notag
\end{align}
with
\begin{equation}
    \Delta \tilde{\textbf{x}}_2(z) = \Delta \tilde{\textbf{y}}_2(z) = \begin{bmatrix}
        -\kappa_{z,2}^2 & 0 & \kappa_{z,2} \tau_{z,2} e^{-i k_y} \\
        0 & 0 & 0 \\
         \kappa_{z,2} \tau_{z,2} e^{i k_y} & 0 & \kappa_{z,2}^2 \\      
    \end{bmatrix}    
    \label{eq:displacement_op_dx2}
\end{equation}
The displacement of the Wannier center of Floquet mode $|\Phi_n\rangle$ in step 2 is then
\begin{align}
    \Delta x_{\mathrm{c},n}(z) & = \frac{1}{4\pi^2}\int_{BZ} \langle\Phi_n | \Delta\tilde{\textbf{x}}_2(z) |\Phi_n \rangle d^2\textbf{k}\notag \\                  
    & = -\frac{1}{4\pi^2}\int_{BZ} \frac{2}{3}\kappa_{z,2}\tau_{z,2} \cos(k_y - \phi_n) d^2\textbf{k} = 0
\end{align}
and similarly $\Delta y_{\mathrm{c},n}(z) = 0$.

In step 3, we have $\mathcal{U}_C(\textbf{k},z) = \mathcal{U}_3(\textbf{k},z) \mathcal{U}_2(\textbf{k},L/2) \mathcal{U}_1(\textbf{k},L/4)$.  The displacement operators can be expressed as
\begin{align}
    \Delta \tilde{\textbf{x}}(z) & = \Delta \tilde{\textbf{x}}_1(L/4) + \Delta\tilde{\textbf{x}}_2(L/2) + \Delta\tilde{\textbf{x}}_3(z)\notag \\
    \Delta \tilde{\textbf{y}}(z) & = \Delta\tilde{\textbf{y}}_2(L/2) + \Delta\tilde{\textbf{y}}_3(z) \notag
\end{align}
with
\begin{align}
    \Delta \tilde{\textbf{x}}_3(z)  & = i\mathcal{U}_1^{\dagger}\mathcal{U}_2^{\dagger}(\mathcal{U}_3^{\dagger}
    \partial_{k_x} \mathcal{U}_3) \mathcal{U}_2 \mathcal{U}_1 \notag \\
    & = \begin{bmatrix}
        0 & 0 & 0 \\
        0 & \kappa_{z,3}^2 & -i\kappa_{z,3} \tau_{z,3} e^{-i k_x} \\
        0 & i \kappa_{z,3} \tau_{z,3} e^{i k_x} & -\kappa_{z,3}^2 \\      
    \end{bmatrix}   
    \label{eq:displacement_op_dx3}
\end{align}
and $\Delta \tilde{\textbf{y}}_3 = \Delta \tilde{\textbf{x}}_3$. The displacement of the Wannier center of Floquet mode $|\Phi_n\rangle$ in step 3 can be computed similar to Eq.(\ref{eq:displacement_step1}) to get $\Delta x_{\mathrm{c},n}(z) = 0$ and $\Delta y_{\mathrm{c},n}(z) = 0$.  

Finally, in step 4, we have $\mathcal{U}_C(\textbf{k},z) = \mathcal{U}_4(\textbf{k},z) \mathcal{U}_3 (\textbf{k},3L/4)\mathcal{U}_2(\textbf{k},L/2) \mathcal{U}_1(\textbf{k},L/4)$.  The displacement operators can be expressed as
\begin{align}
    \Delta \tilde{\textbf{x}}(z) & = \Delta \tilde{\textbf{x}}_1(L/4) + \Delta\tilde{\textbf{x}}_2(L/2) + \Delta\tilde{\textbf{x}}_3(3L/4)\notag \\
    \Delta \tilde{\textbf{y}}(z) & = \Delta\tilde{\textbf{y}}_2(L/2) + \Delta\tilde{\textbf{y}}_3(3L/4) + \Delta\tilde{\textbf{y}}_4(z)\notag
\end{align}
with 
\begin{align}
    \Delta \tilde{\textbf{y}}_4(z) & = i\mathcal{U}_1^{\dagger}\mathcal{U}_2^{\dagger}\mathcal{U}_3^{\dagger}
    (\mathcal{U}_4^{\dagger}\partial_{k_y} \mathcal{U}_4) \mathcal{U}_3 \mathcal{U}_2 \mathcal{U}_1 \notag \\
    & = \begin{bmatrix}
        \kappa_{z,4}^2 & -i\kappa_{z,4} \tau_{z,4} e^{-i k_y} & 0\\
        i \kappa_{z,4} \tau_{z,4} e^{i k_y} & -\kappa_{z,4}^2 & 0 \\  
        0 & 0 & 0 \\
    \end{bmatrix}    
\end{align}
The displacement of the Wannier center of Floquet mode $|\Phi_n\rangle$ in step 4 can also be shown to be $\Delta x_{\mathrm{c},n}(z) = 0$ and $\Delta y_{\mathrm{c},n}(z) = 0$.  Thus the Wannier centers of the three flat-band modes exhibit no motion in $x$ or $y$ over the entire evolution cycle.

\subsection{Helicity of the flat-band modes}

The helicity of the ABF lattice can be computed by defining the helicity operator $\tilde{\textbf{h}} = -3(\Delta\tilde{\textbf{x}}\partial_z\Delta\tilde{\textbf{y}} -
\Delta\tilde{\textbf{y}}\partial_z\Delta\tilde{\textbf{x}})$, so that the helicity of mode $|\Phi_n\rangle$ is then
\begin{equation}
    \mathcal{H}_n = \frac{1}{4\pi^2} \int_0^L dz \int_{BZ} \langle \Phi_n | \tilde{\textbf{h}} |\Phi_n \rangle d^2\textbf{k}
\end{equation}
In step 1, since $\Delta\tilde{\textbf{y}}(z) = 0$, we get $\tilde{\textbf{h}}_1 = 0$ so that the contribution to the helicity in step 1 is zero, $\mathcal{H}_n(1) = 0$.  In step 2, we have
\begin{align}
    \tilde{\textbf{h}}_2 & = -3 \big\{  \big[ \Delta\tilde{\textbf{x}}_1(L/4) + \Delta\tilde{\textbf{x}}_2(z) \big] \partial_z \Delta\tilde{\textbf{y}}_2(z) - \Delta\tilde{\textbf{y}}_2(z) \partial_z\Delta\tilde{\textbf{x}}_2(z) \big \} \notag \\
    & = -3\Delta\tilde{\textbf{x}}_1(L/4) \partial_z \Delta\tilde{\textbf{y}}_2(z)
\end{align}
where we have used the fact that $\Delta\tilde{\textbf{x}}_2 = \Delta\tilde{\textbf{y}}_2$.  Using the matrices for $\Delta\tilde{\textbf{x}}_1$ and $\Delta\tilde{\textbf{y}}_2$ in Eqs.(\ref{eq:displacement_op_dx1}) and (\ref{eq:displacement_op_dx2}), we obtain
\begin{equation}
    \tilde{\textbf{h}}_2 = 3\begin{bmatrix}
        \partial_z(\kappa_{z,2}^2) & 0 & -\partial_z(\kappa_{z,2}\tau_{z,2})e^{-ik_y} \\
        0 & 0 & 0 \\
        0 & 0 & 0
    \end{bmatrix}
\end{equation}
The contribution to the helicity of mode $|\Phi_n\rangle$ in step 2 is
\begin{align}
    \mathcal{H}_n(2) & = \frac{1}{4\pi^2} \int_{L/4}^{L/2} dz \int_{BZ} \langle \Phi_n | \tilde{\textbf{h}}_2 |\Phi_n \rangle d^2\textbf{k} \notag \\
    & = \frac{1}{4\pi^2} \int_{L/4}^{L/2} dz \int_{BZ} [\partial_z(\kappa_{z,2}^2) + \partial_z(\kappa_{z,2}\tau_{z,2})e^{i\phi_n} e^{-ik_y}] d^2\textbf{k} \notag \\
    & = \int_{L/4}^{L/2} \partial_z(\kappa_{z,2}^2)  dz = 1\notag
\end{align}
In step 3, we have
\begin{align}
    \tilde{\textbf{h}}_3 & = -3 \big\{  \big[ \Delta\tilde{\textbf{x}}_1(L/4) + \Delta\tilde{\textbf{x}}_2(L/2) + \Delta\tilde{\textbf{x}}_3(z) \big] \partial_z \Delta\tilde{\textbf{y}}_3(z) - \big[\Delta\tilde{\textbf{y}}_2(L/2) + \Delta\tilde{\textbf{y}}_3(z)\big]
    \partial_z\Delta\tilde{\textbf{x}}_3(z)\big\} \notag \\
    & = -3\Delta\tilde{\textbf{x}}_1(L/4) \partial_z \Delta\tilde{\textbf{y}}_3(z)
\end{align}
where we have also made use of the fact that $\Delta\tilde{\textbf{x}}_3 = \Delta\tilde{\textbf{y}}_3$. Using the matrices for $\Delta\tilde{\textbf{x}}_1$ and $\Delta\tilde{\textbf{y}}_3$ in Eqs.(\ref{eq:displacement_op_dx1}) and (\ref{eq:displacement_op_dx3}), we obtain
\begin{equation}
    \tilde{\textbf{h}}_3 = 3\begin{bmatrix}
        0 & 0 & 0 \\
        0 & \partial_z(\kappa_{z,3}^2) &  -i\partial_z(\kappa_{z,3}\tau_{z,3})e^{-ik_x} \\
        0 & 0 & 0 \\
    \end{bmatrix}
\end{equation}
The contribution to the helicity of mode $|\Phi_n\rangle$ in step 3 is
\begin{align}
    \mathcal{H}_n(3) & = \frac{1}{4\pi^2} \int_{L/2}^{3L/4} dz \int_{BZ} \langle \Phi_n | \tilde{\textbf{h}}_3 |\Phi_n \rangle d^2\textbf{k} \notag \\
    & = \frac{1}{4\pi^2} \int_{L/2}^{3L/4} dz \int_{BZ} [\partial_z(\kappa_{z,3}^2) - \partial_z(\kappa_{z,3}\tau_{z,3})e^{-i\phi_n} e^{-ik_x}] d^2\textbf{k} \notag \\
    & = 1
\end{align}
Finally, in step 4, we have
\begin{equation}
    \tilde{\textbf{h}}_4 = -3\big[\Delta \tilde{\textbf{x}}_1(L/4) + \Delta \tilde{\textbf{x}}_2(L/2) + \Delta \tilde{\textbf{x}}_3(3L/4)\big]\partial_z\Delta\tilde{\textbf{y}}_4
\end{equation}
Using Eqs. (\ref{eq:displacement_op_dx1}), (\ref{eq:displacement_op_dx2}) and (\ref{eq:displacement_op_dx3}), we obtain $\Delta \tilde{\textbf{x}}_1(L/4) + \Delta \tilde{\textbf{x}}_2(L/2) + \Delta \tilde{\textbf{x}}_3(3L/4) = 0$, so that $\tilde{\textbf{h}}_4 = 0$.  Thus the contribution to the helicity in step 4 is $\mathcal{H}_n(4) = 0$.  The helicity of mode $|\Phi_n\rangle$ over one cycle is 
\begin{equation}
    \mathcal{H}_n =  \mathcal{H}_n(1) +  \mathcal{H}_n(2) +  \mathcal{H}_n(3) +  \mathcal{H}_n(4) = 2
\end{equation}
yielding the total helicity of the ABF lattice as $\mathcal{H} = \sum_n \mathcal{H}_n = 6$.

\subsection{Trajectories of the basis components of the flat-band modes}

To determine the evolution of each basis component of the Floquet mode $|\Phi_n\rangle$ in Eq.(\ref{eq:flat_band_modes}), we express the mode in the site basis as
\begin{equation}
    |\Phi_n\rangle = \tfrac{1}{\sqrt{3}}\big(-|A_{\textbf{k}}\rangle + ie^{-i\phi_n}|B_{\textbf{k}}\rangle
    + e^{i\phi_n}|C_{\textbf{k}}\rangle \big ) = \sum_{m} \alpha_{n,m}|m_{\textbf{k}}\rangle
\end{equation}
where $|m_{\textbf{k}}\rangle$ represents field in waveguide $m \in \{A, B, C\}$ and $\alpha_{n,m} = \langle m_{\textbf{k}} | \Phi_n \rangle$.  Starting at $z = 0$, each component of the Floquet mode evolves as $|\psi_m(\textbf{k},z)\rangle = \alpha_{n,m} \mathcal{U}(\textbf{k},z) |m_{\textbf{k}}\rangle$. For instance, the component $|\psi_A\rangle$ evolving from site $A$ during step 1 is
\begin{align}
    |\psi_A (\textbf{k},z)\rangle  & = - \tfrac{1}{\sqrt{3}}e^{i\beta_n z}\mathcal{U}_1(\textbf{k},z) 
    | A_{\textbf{k}} \rangle \notag \\
    & = -\tfrac{1}{\sqrt{3}}e^{i\beta_n z}\big ( \tau_{z,1} |A_{\textbf{k}}\rangle + i\kappa_{z,1} e^{-ik_x} |B_{\textbf{k}}\rangle \big ) 
    \label{eq:WA_step1}
\end{align}
In step 2, it evolves as 
\begin{align}
    |\psi_A (\textbf{k},z)\rangle  & = - \tfrac{1}{\sqrt{3}}e^{i\beta_n z}\mathcal{U}_2(\textbf{k},z) 
    \mathcal{U}_1(\textbf{k},L/4) | A_{\textbf{k}} \rangle \notag \\
    & = -\tfrac{1}{\sqrt{3}}e^{i\beta_n z}\big ( i\tau_{z,2} e^{-ik_x}|B_{\textbf{k}}\rangle - \kappa_{z,2} e^{ik_y} |C_{\textbf{k}}\rangle \big ) 
\end{align}
and in steps 3 and 4, 
\begin{equation}
    |\psi_A(\textbf{k},z)\rangle = \frac{1}{\sqrt{3}}  \begin{cases}
    e^{ik_y}|C_{\textbf{k}}\rangle & L/2 \leq z < 3L/4 \\
    \tau_{z,4}e^{ik_y}|C_{\textbf{k}}\rangle - i\kappa_{z,4}|B_{\textbf{k}}\rangle & 3L/4 \leq z < L \\
    \end{cases}
\end{equation}
The components $|\psi_B\rangle$ and $|\psi_C\rangle$ are similarly computed to give
\begin{equation}
    |\psi_B(\textbf{k},z)\rangle = \frac{i}{\sqrt{3}} e^{-i\phi_n} \begin{cases}
    \tau_{z,1} |B_{\textbf{k}}\rangle +  i\kappa_{z,1} e^{ik_x}|A_{\textbf{k}}\rangle & 0 \leq z < L/4 \\
    ie^{ik_x}|A_{\textbf{k}}\rangle & L/4 \leq z < L/2 \\
    i\tau_{z,3}e^{ik_x}|A_{\textbf{k}}\rangle - i\kappa_{z,3}e^{-ik_y}|B_{\textbf{k}}\rangle & L/2 \leq z < 3L/4 \\
    -\tau_{z,4} e^{-ik_y}|B_{\textbf{k}}\rangle - i\kappa_{z,4} |C_{\textbf{k}}\rangle & 3L/4 \leq z < L
    \end{cases}
\end{equation}
\begin{equation}
    |\psi_C(\textbf{k},z)\rangle = \frac{1}{\sqrt{3}} e^{i\phi_n} \begin{cases}
    |C_{\textbf{k}}\rangle & 0 \leq z < L/4 \\
    \tau_{z,2}|C_{\textbf{k}}\rangle + i\kappa_{z,2} e^{-i(k_x+k_y)}|B_{\textbf{k}}\rangle & L/4 \leq z < L/2 \\
    i\tau_{z,3}e^{-i(k_x+k_y)}|B_{\textbf{k}}\rangle - \kappa_{z,3}|A_{\textbf{k}}\rangle & L/2 \leq z < 3L/4 \\
    -|A_{\textbf{k}}\rangle & 3L/4 \leq z < L
    \end{cases}
\end{equation}
Using Eq.(26) in the main text, we can compute the Wannier functions of the three components in the home unit cell $(0,0)$.  The resulting normalized functions are as follows:
\begin{equation}
    |w_A(x,y,z)\rangle = \begin{cases}
    \tau_{z,1}|A_{0,0}\rangle + i\kappa_{z,1}|B_{1,0}\rangle & 0 \leq z < L/4 \\
    i\tau_{z,2}|B_{1,0}\rangle - \kappa_{z,2}|C_{0,-1}\rangle & L/4 \leq z < L/2 \\
    -|C_{0,-1}\rangle & L/2 \leq z < 3L/4 \\
    -\tau_{z,4}|C_{0,-1}\rangle + i\kappa_{z,4}|B_{0,0}\rangle & 3L/4 \leq z < L
    \end{cases}
    \label{eq:WA}
\end{equation}
\begin{equation}
    |w_B(x,y,z)\rangle = \begin{cases}
    \tau_{z,1}|B_{0,0}\rangle + i\kappa_{z,1}|A_{-1,0}\rangle & 0 \leq z < L/4 \\
    i|A_{-1,0}\rangle  & L/4 \leq z < L/2 \\
    i\tau_{z,3}|A_{-1,0}\rangle - \kappa_{z,3}|B_{0,1}\rangle & L/2 \leq z < 3L/4 \\
    -\tau_{z,4}|B_{0,1}\rangle - i\kappa_{z,4}|C_{0,0}\rangle & 3L/4 \leq z < L
    \end{cases}
    \label{eq:WB}
\end{equation}
\begin{equation}
    |w_C(x,y,z)\rangle = \begin{cases}
    |C_{0,0}\rangle  & 0 \leq z < L/4 \\
    \tau_{z,2}|C_{0,0}\rangle + i\kappa_{z,2}|B_{1,1}\rangle & L/4 \leq z < L/2 \\
    i\tau_{z,3}|B_{1,1}\rangle - \kappa_{z,3}|A_{0,0}\rangle & L/2 \leq z < 3L/4 \\
    -|A_{0,0}\rangle & 3L/4 \leq z < L
    \end{cases}
    \label{eq:WC}
\end{equation}
The field distributions of the three component Wannier functions in Eqs.(\ref{eq:WA})-(\ref{eq:WC}) over one cycle are shown in Fig. 6(a) in the main text.

The trajectory of the center of each component Wannier function can be calculated using the position operator $\tilde{\textbf{r}}$, which gives the $(x, y)$ coordinates of waveguide $M \in \{A, B, C\}$ when acting on the corresponding basis state $|M_{m,n}\rangle$.  For the home unit cell $(0,0)$ the coordinates of waveguides $A, B, C$ are assumed to be $[1/2, 0]^\mathrm{T}, [-1/2, -1/2]^\mathrm{T}, [0, 1/2]^\mathrm{T}$, respectively, as depicted in Fig. 6(b).  The center of the Wannier function $|w_A(x, y, z)\rangle$ during step 1 (Eq.(\ref{eq:WA})) can then be calculated as
\begin{align}
    \textbf{r}_A(z) & = \langle w_A(x, y, z) | \tilde{\textbf{r}} | w_A(x, y, z) \rangle \notag \\
    & = \tau_{z,1}^2 \langle A_{0,0} | \tilde{\textbf{r}} | A_{0,0} \rangle + \kappa_{z,1}^2
     \langle B_{1,0} | \tilde{\textbf{r}} | B_{1,0} \rangle \notag \\
    & = \tau_{z,1}^2 \begin{bmatrix} 1/2 \\ 0 \end{bmatrix} + \kappa_{z,1}^2 \begin{bmatrix} 1/2 \\ -1/2 \end{bmatrix} = \begin{bmatrix} 1/2 \\ -\kappa_{z,1}^2/2 \end{bmatrix} \notag
\end{align}
The complete trajectories of the three components of the flat-band mode $|\Phi_n\rangle$ are given in Eq. (28) in the main text.

\section{Relation between the helicity and the twist number}

To connect the non-Abelian helicity in Eq.(12) to the twist number in Eq.(14) in the main text, we first show that $\textbf{A} \cdot \grad \cross \textbf{A} = i\textbf{A} \cdot \textbf{A} \cross \textbf{A}$.  Writing $\textbf{A} \cdot \grad \cross \textbf{A}$ in component form, we get
\begin{equation}
    \textbf{A} \cdot \grad \cross \textbf{A} = -\mathcal{T} + \mathcal{F}_1 + \mathcal{F}_2
    \label{eq:H_op}
\end{equation}
where
\begin{align}
    \mathcal{T} &= A_{k_x}\partial_z A_{k_y} - A_{k_y}\partial_z A_{k_x} \label{eq:T} \\
    \mathcal{F}_1 &= A_z(\partial_{k_x} A_{k_y} - \partial_{k_y} A_{k_x}) \label{eq:F1} \\
    \mathcal{F}_2 &= A_{k_x} \partial_{k_y} A_z - A_{k_y} \partial_{k_x} A_z \label{eq:F2}
\end{align}
The matrices $A_{k_x}$, $A_{k_y}$ and $A_z$ can be related to the evolution operator $\mathcal{U}$ as [31]
\begin{equation}
    A_{k_x} = i \mathcal{U}^{\dagger} \partial_{k_x} \mathcal{U}; A_{k_y} = i \mathcal{U}^{\dagger} \partial_{k_y} \mathcal{U}; A_z = i\mathcal{U}^{\dagger} \partial_z \mathcal{U}
    \label{eq:A_to_U}
\end{equation}
Using these relations, we can express the terms in the brackets in Eq.(\ref{eq:F1}) as follows:
\begin{align}
    \partial_{k_x} A_{k_y} - \partial_{k_y} A_{k_x} &= i\partial_{k_x}(\mathcal{U}^{\dagger} \partial_{k_y} \mathcal{U}) - i \partial_{k_y}(\mathcal{U}^{\dagger} \partial_{k_x} \mathcal{U}) \notag \\
    &= i\partial_{k_x}\mathcal{U}^{\dagger}\partial_{k_y}\mathcal{U} - i \partial_{k_y}\mathcal{U}^{\dagger}\partial_{k_x} \mathcal{U} \label{eq:AkxAky}
\end{align}
Since $\partial_{k}(\mathcal{U}^{\dagger}\mathcal{U}) = \partial_{k}\mathcal{U}^{\dagger}\mathcal{U} + \mathcal{U}^{\dagger}\partial_{k}\mathcal{U} = 0 $ for $k \in \{k_x, k_y\}$, we have $\partial_{k}\mathcal{U}^{\dagger} = -\mathcal{U}^{\dagger} \partial_k \mathcal{U} \mathcal{U}^{\dagger}$.  Using this relationship in Eq.(\ref{eq:AkxAky}), we obtain
\begin{align}
    \partial_{k_x} A_{k_y} - \partial_{k_y} A_{k_x} &= -i (\mathcal{U}^{\dagger} \partial_{k_x} \mathcal{U}) 
    (\mathcal{U}^{\dagger} \partial_{k_y} \mathcal{U}) + i (\mathcal{U}^{\dagger} \partial_{k_y} \mathcal{U})
    (\mathcal{U}^{\dagger} \partial_{k_x} \mathcal{U})  \notag \\
    &= i A_{k_x} A_{k_y} - i A_{k_y} A_{k_x}
\end{align}
Substituting the above result into Eq.(\ref{eq:F1}) gives
\begin{equation}
    \mathcal{F}_1 = i (A_z A_{k_x} A_{k_y} - A_z A_{k_y} A_{k_x})
    \label{eq:F1b}
\end{equation}
To evaluate the term $\mathcal{F}_2$, we first compute the partial derivatives $\partial_k A_z$, $k \in \{k_x, k_y\}$. Since $\partial_z \mathcal{U} = i H \mathcal{U}$, we can express $A_z$ as $A_z = i\mathcal{U}^{\dagger} \partial_z \mathcal{U} = -\mathcal{U}^{\dagger} H \mathcal{U}$. We thus have
\begin{align}
    \partial_k A_z & = -\partial_k(\mathcal{U}^{\dagger} H \mathcal{U}) = -\partial_k \mathcal{U}^{\dagger}H \mathcal{U}
    - \mathcal{U}^{\dagger} \partial_k H \mathcal{U} - \mathcal{U}^{\dagger} H \partial_k \mathcal{U} \notag \\
     & = (\mathcal{U}^{\dagger} \partial_k \mathcal{U})( \mathcal{U}^{\dagger} H \mathcal{U}) - \mathcal{U}^{\dagger} \partial_k H \mathcal{U} - (\mathcal{U}^{\dagger} H \mathcal{U})( \mathcal{U}^{\dagger} \partial_k \mathcal{U}) \notag \\
   &=  \partial_z A_k + i(A_k A_z - A_z A_k)
   \label{eq:partial_k_A_z}
\end{align}
In the above expression we have made use of the relationship $\partial_z A_k = -\mathcal{U}^{\dagger} \partial_k H \mathcal{U}$, which can be shown as follows:
\begin{align}
    \partial_z A_k & = \partial_z( i\mathcal{U}^{\dagger} \partial_k \mathcal{U}) = i\partial_z \mathcal{U}^{\dagger} \partial_k \mathcal{U} + i \mathcal{U}^{\dagger}
    \partial_k (\partial_z  \mathcal{U}) \notag \\
    & = (H \mathcal{U})^{\dagger} \partial_k \mathcal{U} -\mathcal{U}^{\dagger} \partial_k( H \mathcal{U}) = -\mathcal{U}^{\dagger} \partial_k H \mathcal{U}  \notag
\end{align}
where we have made use of the fact that $H = H^{\dagger}$.  
Substituting Eq.(\ref{eq:partial_k_A_z}) into Eq.(\ref{eq:F2}), we get
\begin{align}
    \mathcal{F}_2 &= A_{k_x} (\partial_z A_{k_y} + i A_{k_y} A_z  - i A_z A_{k_y} ) - A_{k_y} ( \partial_z A_{k_x} + i A_{k_x} A_z  - i A_z A_{k_x} ) \notag \\
    &= \mathcal{T} + i(A_{k_x} A_{k_y} A_z - A_{k_x} A_z A_{k_y} 
    + A_{k_y} A_z A_{k_x} - A_{k_y} A_{k_x} A_z)
    \label{eq:F2b}
\end{align}
Upon substituting Eqs.(\ref{eq:F1b}) and (\ref{eq:F2b}) into Eq.(\ref{eq:H_op}), we obtain
\begin{equation}
    \textbf{A} \cdot \grad \cross \textbf{A} = i\textbf{A} \cdot \textbf{A} \cross \textbf{A}
    \label{eq:H_op2}
\end{equation}

To connect the helicity to the twist number, we take the trace of Eq.(\ref{eq:H_op2}) and make use of the cyclical property of the trace to obtain
\begin{align}
    \mathrm{Tr}(\textbf{A} \cdot \grad \cross \textbf{A}) &= i\mathrm{Tr}(\textbf{A} \cdot \textbf{A} \cross \textbf{A}) = 3i \mathrm{Tr}\{A_z (A_{k_x} A_{k_y} - A_{k_y} A_{k_x})\}
\end{align}
Next using the relationships in Eq.(\ref{eq:A_to_U}), we get
\begin{align}
    \mathrm{Tr}(\textbf{A} \cdot \grad \cross \textbf{A})
    &= 3i \mathrm{Tr}\{(\mathcal{U}^{\dagger} H \mathcal{U})[( \mathcal{U}^{\dagger} \partial_{k_x} \mathcal{U} )
    (\mathcal{U}^{\dagger} \partial_{k_y} \mathcal{U}) - 
     (\mathcal{U}^{\dagger} \partial_{k_y} \mathcal{U})( \mathcal{U}^{\dagger} \partial_{k_x} \mathcal{U})]\} \notag \\
     & = 3i \mathrm{Tr}\{\mathcal{U}^{\dagger} H \partial_{k_x} \mathcal{U}(\mathcal{U}^{\dagger} \partial_{k_y} \mathcal{U}) - \mathcal{U}^{\dagger} H \partial_{k_y}\mathcal{U} (\mathcal{U}^{\dagger} \partial_{k_x} \mathcal{U})\}
\end{align}  
Using the relation $\mathcal{U}^{\dagger} \partial_k \mathcal{U} = -\partial_k \mathcal{U}^{\dagger} \mathcal{U}$, we can simplify the above expression to
\begin{align}
    \mathrm{Tr}(\textbf{A} \cdot \grad \cross \textbf{A}) &= 3i \mathrm{Tr}\{-\mathcal{U}^{\dagger} H \partial_{k_x} \mathcal{U} (\partial_{k_y} \mathcal{U}^{\dagger} \mathcal{U}) + \mathcal{U}^{\dagger} H \partial_{k_y}\mathcal{U} (\partial_{k_x} \mathcal{U}^{\dagger} \mathcal{U})\} \notag \\
    &= -3i \mathrm{Tr}\{H \partial_{k_x} \mathcal{U} \partial_{k_y} \mathcal{U}^{\dagger} - H \partial_{k_y}\mathcal{U} \partial_{k_x} \mathcal{U}^{\dagger} \}
    \label{eq:T1}
\end{align}
From
\begin{align*}
    \partial_{k_x} (H \mathcal{U} \partial_{k_y} \mathcal{U}^{\dagger}) &= \partial_{k_x} H \mathcal{U}
    \partial_{k_y} \mathcal{U}^{\dagger} + H \partial_{k_x} \mathcal{U} \partial_{k_y} \mathcal{U}^{\dagger} 
    + H \mathcal{U} \partial_{k_x}\partial_{k_y} \mathcal{U}^{\dagger} \\
    \partial_{k_y} (H \mathcal{U} \partial_{k_x} \mathcal{U}^{\dagger}) &= \partial_{k_y} H \mathcal{U}
    \partial_{k_x} \mathcal{U}^{\dagger} + H \partial_{k_y} \mathcal{U} \partial_{k_x} \mathcal{U}^{\dagger} 
    + H \mathcal{U} \partial_{k_y}\partial_{k_x} \mathcal{U}^{\dagger}
\end{align*}
subtracting the two expressions gives
\begin{align*}
    H \partial_{k_x} \mathcal{U} \partial_{k_y} \mathcal{U}^{\dagger} - H \partial_{k_y}\mathcal{U} \partial_{k_x} \mathcal{U}^{\dagger} = & \partial_{k_x} (H \mathcal{U} \partial_{k_y} \mathcal{U}^{\dagger}) - 
    \partial_{k_y} (H \mathcal{U} \partial_{k_x} \mathcal{U}^{\dagger}) \\
    & - (\partial_{k_x} H \mathcal{U} \partial_{k_y} \mathcal{U}^{\dagger}  
    - \partial_{k_y} H \mathcal{U} \partial_{k_x} \mathcal{U}^{\dagger})
\end{align*}
Substituting the above result into Eq.(\ref{eq:T1}) and integrating over a BZ, we obtain
\begin{equation}
    \frac{1}{4\pi^2}\int_{BZ} \mathrm{Tr}(\textbf{A} \cdot \grad \cross \textbf{A}) d^2\textbf{k} = \frac{3i}{4\pi^2} \int_{BZ} \mathrm{Tr} (
    \partial_{k_x} H \mathcal{U} \partial_{k_y} \mathcal{U}^{\dagger} 
    - \partial_{k_y} H \mathcal{U} \partial_{k_x} \mathcal{U}^{\dagger} ) d^2\textbf{k}
    \label{eq:T2}
\end{equation}
where the integrals over the terms $\partial_{k_x} (H \mathcal{U} \partial_{k_y} \mathcal{U}^{\dagger})$ and $\partial_{k_y} (H \mathcal{U} \partial_{k_x} \mathcal{U}^{\dagger})$ vanish due to the periodic conditions on the BZ boundaries.  Substituting $\partial_k \mathcal{U}^{\dagger} = -\mathcal{U}^{\dagger} \partial_k \mathcal{U} \mathcal{U} ^{\dagger}$ into Eq.(\ref{eq:T2}), we get
\begin{align*}
    \frac{1}{4\pi^2}\int_{BZ} \mathrm{Tr}(\textbf{A} \cdot \grad \cross \textbf{A}) d^2\textbf{k} 
    &= -\frac{3i}{4\pi^2} \int_{BZ} \mathrm{Tr} \{
    (\partial_{k_x} H \mathcal{U})( \mathcal{U}^{\dagger} \partial_{k_y} \mathcal{U} \mathcal{U} ^{\dagger} )
    - (\partial_{k_y} H \mathcal{U})( \mathcal{U}^{\dagger} \partial_{k_x} \mathcal{U} \mathcal{U} ^{\dagger}) \} d^2\textbf{k}\\
    &= -\frac{3i}{4\pi^2} \int_{BZ} \mathrm{Tr} \{
    (\mathcal{U} ^{\dagger}\partial_{k_x} H \mathcal{U})( \mathcal{U}^{\dagger} \partial_{k_y} \mathcal{U}  )
    - (\mathcal{U} ^{\dagger}\partial_{k_y} H \mathcal{U})( \mathcal{U}^{\dagger} \partial_{k_x} \mathcal{U}) \} d^2\textbf{k}\\
    &=
    \frac{3}{4\pi^2} \int_{BZ} \mathrm{Tr} (\partial_z A_{k_x}A_{k_y} - \partial_z A_{k_y} A_{k_x}) d^2\textbf{k} \\
    &=
    -\frac{3}{4\pi^2} \int_{BZ} \mathrm{Tr} (A_{k_x} \partial_z A_{k_y} - A_{k_y} \partial_z A_{k_x} ) d^2\textbf{k} \\
    &= -\frac{3}{4\pi^2} \int_{BZ} \mathrm{Tr} (\tilde{\textbf{r}} \cross \partial_z \tilde{\textbf{r}} \cdot \hat{\textbf{z}}) d^2\textbf{k}
\end{align*}
Integrating over an evolution period $L$ gives the helicity in terms of the twist operator in Eq. (14).

\end{document}